\documentclass[journal=jpclcd,manuscript=letter]{achemso}

\usepackage[utf8]{inputenc}
\usepackage{chemformula} % Formula subscripts using \ch{}
\usepackage{chemfig}
\usepackage[T1]{fontenc} % Use modern font encodings
\usepackage{graphicx}% Include figure files
\usepackage{dcolumn}% Align table columns on decimal point
\usepackage{siunitx}
\DeclareSIUnit\angstrom{\text{Å}}
\usepackage{amsmath}
\usepackage{multirow}
\usepackage[ruled]{algorithm2e}
\usepackage{hyperref}
\hypersetup{
    colorlinks=true,
    linkcolor=cyan,
    filecolor=magenta,      
    urlcolor=cyan,
    citecolor=cyan,
    }

\DeclareMathVersion{sans}
\SetSymbolFont{operators}{sans}{OT1}{cmbr}{m}{n}
\SetSymbolFont{letters}{sans}{OML}{cmbr}{m}{it}
\SetSymbolFont{symbols}{sans}{OMS}{cmbr}{m}{n}
\SetMathAlphabet{\mathit}{sans}{OT1}{cmbr}{m}{sl}
\SetMathAlphabet{\mathbf}{sans}{OT1}{cmbr}{bx}{n}
\SetMathAlphabet{\mathtt}{sans}{OT1}{cmtl}{m}{n}
\SetSymbolFont{largesymbols}{sans}{OMX}{cmbr}{m}{n}

\mathversion{sans}

\SectionNumbersOn

\newcommand\dd{\mathop{}\!\mathrm{d}} 
\newcommand\mvec[1]{\underline{\mathbf{#1}}}

\title{Including photoexcitation explicitly in trajectory-based nonadiabatic dynamics at no cost}

\author{Ji\v{r}\'{i} Jano\v{s}}
\email{jiri.janos@vscht.cz}
\affiliation[VSCHT]
{Department of Physical Chemistry, University of Chemistry and Technology, Technická 5, Prague 6, 166 28, Czech Republic}
\alsoaffiliation[UB]
{Centre for Computational Chemistry, School of Chemistry, University of Bristol, Bristol BS8 1TS, United Kingdom}

\author{Petr Slav\'{i}\v{c}ek}
\email{petr.slavicek@vscht.cz}
\affiliation[VSCHT]
{Department of Physical Chemistry, University of Chemistry and Technology, Technická 5, Prague 6, 166 28, Czech Republic}

\author{Basile F. E. Curchod}
\email{basile.curchod@bristol.ac.uk}
\affiliation[UB]
{Centre for Computational Chemistry, School of Chemistry, University of Bristol, Bristol BS8 1TS, United Kingdom}

\begin{document}

\begin{tocentry}
% 5x5cm
\includegraphics[width=\textwidth]{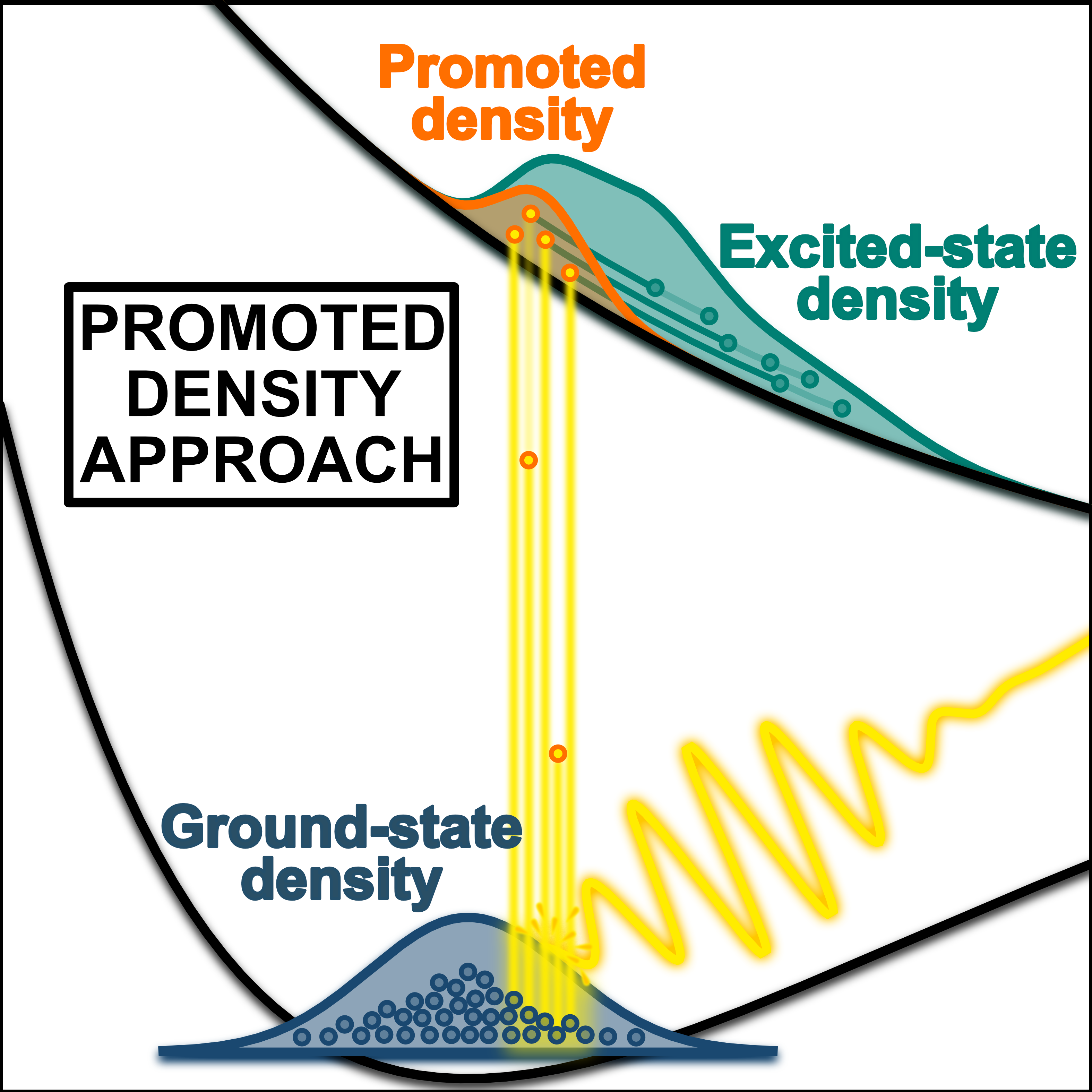}

\end{tocentry}

\renewcommand{\baselinestretch}{1.0}\normalsize

\begin{abstract}

Over the last decades, theoretical photochemistry has produced multiple techniques to simulate the nonadiabatic dynamics of molecules.  Surprisingly, much less effort has been devoted to adequately describing the first step of a photochemical or photophysical process: photoexcitation. Here, we propose a formalism to include the effect of a laser pulse in trajectory-based nonadiabatic dynamics at the level of the initial conditions, with no additional cost. The promoted density approach (PDA) decouples the excitation from the nonadiabatic dynamics by defining a new set of initial conditions, which include an excitation time. PDA with surface hopping leads to nonadiabatic dynamics simulations in excellent agreement with quantum dynamics using an explicit laser pulse and highlights the strong impact of a laser pulse on the resulting photodynamics and the limits of the (sudden) vertical excitation. Combining PDA with trajectory-based nonadiabatic methods is possible for any arbitrary-sized molecules using a code provided in this work. 

\end{abstract}


\renewcommand{\baselinestretch}{1.2}\normalsize

Since the development of femtochemistry, the field of ultrafast spectroscopy has been flourishing, offering over the years a plethora of strategies to investigate the dynamics of photoexcited molecules.\cite{Centurion2022, D1CP05885A, Zhang2022, Rolles2023} Current cutting-edge experiments performed at advanced light sources can probe photochemical reactions with precisely controlled femto/attosecond laser pulses and measure fine details of nuclear and sometimes even electronic evolution.\cite{Nisoli2017} This fast experimental pace has, of course, also greatly stimulated the development of theoretical approaches to describe the excited-state dynamics of molecules beyond the Born-Oppenheimer approximation, also called nonadiabatic molecular dynamics.\cite{persico2014overview,agostini2019different,https://doi.org/10.1002/anie.201916381,gonzalez2020quantum, Crespo-Otero2018}

Over the last decades, the field of nonadiabatic molecular dynamics has provided a large toolbox of methods for describing the evolution of molecules following photoexcitation to an excited electronic state,\cite{Beck2000, Richings2015, Curchod2018} often based on trajectories when the molecule is treated in its full dimensionality.\cite{tully2012perspective, Barbatti2011, Crespo-Otero2018} When combined with high-level electronic-structure techniques to describe the underlying electronic energies and nonadiabatic couplings, nonadiabatic molecular dynamics methods can be used to accurately model the excited-state dynamics of molecules and provide interpretation and guidance to experiments. 

Yet, to describe a photochemical experiment reliably, it is crucial to account for all its steps: excitation of the molecule by a laser pulse (or light), evolution in the (coupled) excited electronic states, and formation of photoproducts (in the excited or ground electronic state). In contrast to the extensive developments in electronic-structure theory and nonadiabatic dynamics, it is striking to realize how primitive is the description of the photoexcitation process in current nonadiabatic simulations.\cite{Suchan2018} As an example, let us consider a recent prediction challenge presented to the nonadiabatic-dynamics community, aiming at simulating the photochemistry of cyclobutanone and predicting the time-resolved ultrafast electron diffraction signal resulting from this dynamics before the experiment is conducted.\cite{predictionchallenge} While the experimentalists provided all the details about the laser pulse they would use for the photoexcitation of cyclobutanone, not a single group incorporated the pulse explicitly, including the authors of the present work. This observation can be rationalized by the associated computational cost of including a laser pulse explicitly in nonadiabatic dynamics simulations\cite{Kim2015, Mignolet2016,Makhov2018} and the fact that including an explicit laser pulse tends to stretch the approximations of mixed quantum/classical methods like fewest-switches surface hopping (FSSH).\cite{bajo2014interplay,Mignolet2019b} However, it is still striking that almost half of the proposed predictions completely ignored the underlying effect of the laser pulse on the photoexcitation, that is, the energy spectrum and temporal spread of the pulse. The remaining predictions used a simple energy window based on the pulse spectrum to select initial conditions for the dynamics, yet with multiple definitions for the window. The pulse duration was also often ignored assuming instantaneous excitation. This recent observation, based on state-of-the-art simulations, reveals the following: \textit{the field of nonadiabatic molecular dynamics critically lacks a standard approach to account for the photoexcitation by a laser pulse properly}. 

In this Letter, we propose to establish a systematic scheme for the explicit inclusion of photoexcitation in nonadiabatic molecular dynamics simulations at the level of the initial conditions, with no additional cost. This strategy is derived from the concept of promoted nuclear density -- and as such coined promoted density approach (PDA) -- and can be straightforwardly applied to trajectory-based nonadiabatic molecular dynamics methods as it simply selects from the traditional initial conditions (nuclear positions and momenta) based on the laser pulse spectral intensity and complements them with an 'excitation time'. The PDA can also be used to justify a proper way of performing energy windowing and convolution in nonadiabatic dynamics. 

PDA finds its origin in the derivation of a time-dependent excited-state nuclear density formula based on first-order perturbation theory, considering a weak field regime for the laser pulse. The first work on this topic was proposed by Martens et al,\cite{Li1996} further extended by Shen and Cina\cite{Shen1999} and later by Meyer and Engel\cite{Meier2002}, all considering a Gaussian laser pulse and constant transition dipole moment. Following their work, Martínez-Mesa and Saalfrank generalized the equation for an arbitrary pulse envelope using the pulse envelope Wigner representation.\cite{Martinez-Mesa2015} PDA builds on all the aforementioned works and encompasses any arbitrary laser pulse, as well as position-dependent transition dipole moments. 

Let us discuss the key steps in the derivation that lead to PDA (a detailed derivation is proposed in the SI). We consider a molecular system with a ground ($g$) and an excited ($e$) electronic state, coupled through a weak interaction defined in the position representation as
\begin{equation}
    \hat{V}_\mathrm{int}(\mvec{R},t) = -\Vec{\mu}_{eg}(\mvec{R})\cdot\Vec{E}_0 E(t) \, ,
\end{equation}
where $\Vec{\mu}_{eg}$ denotes the transition dipole moment depending on nuclear configuration $\mvec{R}$, $\Vec{E}_0$ is the electric field amplitude $E_0$ multiplied by the polarization vector $\Vec{\lambda}$ of the field, and $E(t)$ is a real time-dependent electric field. The excited-state nuclear density for such a system can be expressed using first-order perturbation theory as
\begin{equation}
    \rho_e(\mvec{R},t) = \frac{1}{\hbar^2} \mathrm{e}^{\mathcal{L}_e t} \int_{-\infty}^{\infty} \int_{-\infty}^{\infty}  \mathrm{e}^{\frac{i}{\hbar}\hat{H}_e \tau^\prime} \hat{V}_\mathrm{int}(\mvec{R},\tau^\prime) \mathrm{e}^{-\frac{i}{\hbar}\hat{H}_g \tau^\prime} \rho_g(\mvec{R}) \mathrm{e}^{\frac{i}{\hbar}\hat{H}_g \tau} \hat{V}_\mathrm{int}(\mvec{R},\tau)  \mathrm{e}^{-\frac{i}{\hbar}\hat{H}_e \tau}\dd \tau^\prime  \dd \tau \, ,
    \label{eq:dens_prop}
\end{equation}
where $\mathrm{e}^{\mathcal{L}_e t}$ is the excited-state quantum Liouville propagator defined as $\mathrm{e}^{\mathcal{L}_e t} \rho = \mathrm{e}^{-\frac{i}{\hbar}\hat{H}_e t} \rho \mathrm{e}^{\frac{i}{\hbar}\hat{H}_e t}$ and $\hat{H}_{g/e} = \hat{T} + E^\mathrm{el}_{g/e}(\mvec{R})$ is the time-independent Hamiltonian for either the ground or excited electronic state with $E^\mathrm{el}_{g/e}(\mvec{R})$ standing for the respective electronic potential energy surfaces (and $\hat{T}$ for the nuclear kinetic energy operator). We note that also the Liouvillian and Hamiltonian are position-dependent, but we drop $\mvec{R}$ here to simplify the notation. The ground-state nuclear density, $\rho_g$, is considered to be stationary under the $\hat{H}_g$ Hamiltonian. 

The integrals in Eq.~\eqref{eq:dens_prop} can be disentangled by substituting $\tau = t^\prime + \frac{s}{2}$ and $\tau^\prime = t^\prime - \frac{s}{2}$, leading to 
\begin{align}
    \rho_e(\mvec{R},t) = \frac{1}{\hbar^2} \int_{-\infty}^{\infty} \mathrm{e}^{\mathcal{L}_e (t-t^\prime)} \rho_p(\mvec{R},t^\prime) \dd t^\prime \, ,\label{eq:promden1}
\end{align}
where $\rho_p$ is a promoted nuclear density, defined as 
\begin{align}
    \rho_p(\mvec{R},t^\prime) &=  \int_{-\infty}^{\infty}  \mathrm{e}^{-\frac{i}{\hbar}\hat{H}_e \frac{s}{2}}  \hat{V}_\mathrm{int}\left(\mvec{R},t^\prime - \frac{s}{2}\right) \mathrm{e}^{\frac{i}{\hbar}\hat{H}_g\frac{s}{2}} \rho_g(\mvec{R}) \mathrm{e}^{\frac{i}{\hbar}\hat{H}_g \frac{s}{2}}  \hat{V}_\mathrm{int}\left(\mvec{R},t^\prime + \frac{s}{2}\right) \mathrm{e}^{-\frac{i}{\hbar}\hat{H}_e \frac{s}{2}} \dd s \, .
    \label{eq:rhof1}
\end{align}
Identifying the promoted density above allows us to separate the photoexcitation process (Eq.~\eqref{eq:rhof1}) from the excited-state dynamics (Eq.~\eqref{eq:promden1}). The promoted nuclear density is the central quantity of our derivation and represents the nuclear density promoted to the excited electronic state at time $t^\prime$, which we call the excitation time. As such, $\rho_p$ is proportional to the interaction strength represented as $E_0$. The promoted nuclear density, once propagated in the excited electronic state from time $t^\prime$ to $t$ and integrated over all excitation times $t^\prime$, reconstructs the correct excited-state density at time $t$, $\rho_e(\mvec{R},t)$. 

The expression for the promoted nuclear density, Eq.~\eqref{eq:rhof1}, can be simplified by applying twice the first-order Baker--Campbell--Hausdorff formula to the operators on both sides of the ground-state density $\rho_g$ (see SI for the detailed derivation and analysis of terms neglected), leading to 
\begin{align}
    \rho_p(\mvec{R},t^\prime) &= |\Vec{\mu}_{eg}(\mvec{R})\cdot\Vec{E}_0|^2 \left[\int_{-\infty}^{\infty} E(t^\prime + \frac{s}{2}) E(t^\prime - \frac{s}{2}) \mathrm{e}^{-\frac{i}{\hbar}\Delta E^\mathrm{el}_{eg}(\mvec{R}) s} \dd s\right] \rho_g(\mvec{R}) \notag \\
    &= |\Vec{\mu}_{eg}(\mvec{R})\cdot\Vec{E}_0|^2 \mathcal{W}_E(t^\prime,\Delta E^\mathrm{el}_{eg}(\mvec{R})/\hbar) \rho_g(\mvec{R})  \, ,\label{eq:finalden}
\end{align}
where we have identified the term between squared brackets as the Wigner pulse representation $\mathcal{W}_E$\cite{DIELS20061}, defined as 
\begin{align}
    \mathcal{W}_E(t^\prime,\omega) &= \int_{-\infty}^{\infty} E\left(t^\prime+\frac{s}{2}\right) E^*\left(t^\prime-\frac{s}{2}\right) \mathrm{e}^{-i\omega s} \dd s \, .
    \label{eq:wigner}
\end{align}
The Wigner pulse representation can be viewed as a quasiprobability function for the laser pulse to have a certain pulse frequency at a given time, similar to the concept of a Wigner distribution in quantum mechanics. We will come back to the meaning and properties of $\mathcal{W}_E$ as soon as we finish this derivation. 

Finally, inserting the promoted density into the expression for the excited-state nuclear density  (Eq.~\eqref{eq:promden1}) and taking a classical limit by retaining only the lowest-order term in $\hbar$ in the Wigner transform of the density operator results in the final formula for PDA,

\begin{align}
    \rho^\mathrm{cl}_e(\mvec{R},\mvec{P},t) & = \frac{1}{\hbar^2} \int_{-\infty}^{\infty} \mathrm{e}^{\mathcal{L}^\mathrm{cl}_e (t-t^\prime)} \rho^\mathrm{cl}_p(\mvec{R},\mvec{P},t') \dd t^\prime \notag \\
    &= \frac{1}{\hbar^2} \int_{-\infty}^{\infty} \mathrm{e}^{\mathcal{L}^\mathrm{cl}_e (t-t^\prime)} \left[ |\Vec{\mu}_{eg}(\mvec{R})\cdot\Vec{E}_0|^2 \mathcal{W}_E(t^\prime,\Delta E^\mathrm{el}_{eg}(\mvec{R})/\hbar)  \rho^\mathrm{cl}_g(\mvec{R},\mvec{P}) \right] \dd t^\prime \, ,
    \label{eq:dens_prop3}
\end{align}
where $\mvec{P}$ stands for the nuclear momenta and $\mathcal{L}^\mathrm{cl}$ is the classical Liouvillian. This semiclassical approximation limits the formula to cases where the interference effects in the excited state can be ignored (see SI). 

Eq.~\eqref{eq:dens_prop3} as such is derived to describe the excited-state nuclear density following the interaction with the laser pulse (see SI for a detailed discussion). However, we can extend the reach of Eq.~\eqref{eq:dens_prop3} such that it describes the excited-state density also during the pulse by altering the upper integration limit to $t$ instead of $\infty$,
\begin{align}
    \rho^\mathrm{cl}_e(\mvec{R},\mvec{P},t) &= \frac{1}{\hbar^2} \int_{-\infty}^{t} \mathrm{e}^{\mathcal{L}^\mathrm{cl}_e (t-t^\prime)} \left[ |\Vec{\mu}_{eg}(\mvec{R})\cdot\Vec{E}_0|^2 \mathcal{W}_E(t^\prime,\Delta E^\mathrm{el}_{eg}(\mvec{R})/\hbar)  \rho^\mathrm{cl}_g(\mvec{R},\mvec{P}) \right] \dd t^\prime \, .
    \label{eq:dens_prop3b}
\end{align}
This modification means that only the contributions of the promoted density to $\rho_e$ coming from times before the current time $t$ are included. Any contributions to $\rho_e$ coming from the promoted density at later times $t^\prime>t$ are neglected. This empirical alteration of the upper integration limit affects the overall equation only during the pulse interaction and not for times $t$ after the pulse. This modification allows PDA to describe the excited-state nuclear density \textit{within} the pulse duration while not altering the final excited-state density after the pulse. More details on this procedure (and its numerical validation) are provided in the SI.

\begin{figure}[ht!]
    \centering
    \includegraphics[width=0.9\textwidth]{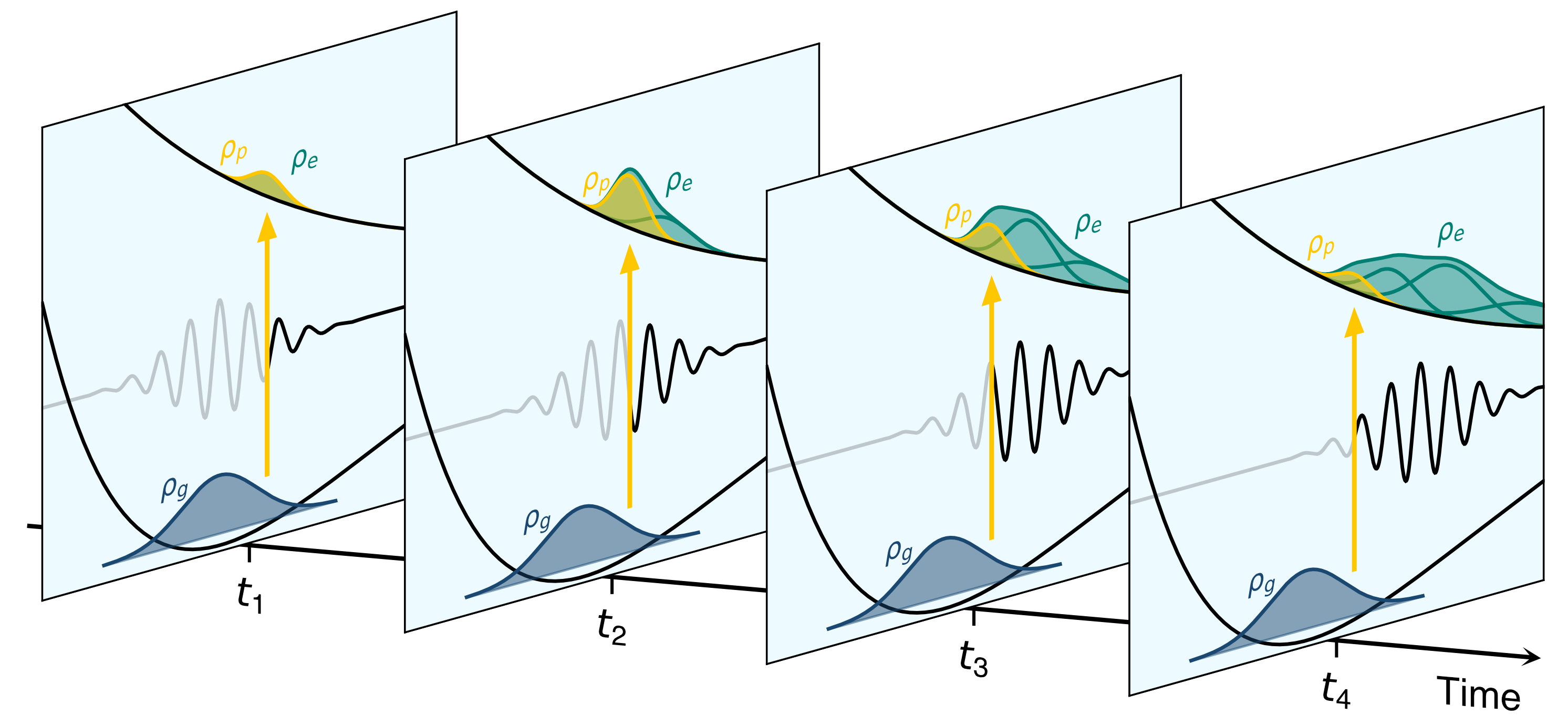}
    \caption{Illustration of photoexcitation and subsequent dynamics of the excited-state nuclear density $\rho_e$ (green) depicted in terms of ground-state nuclear density $\rho_g$ (blue) and promoted nuclear density $\rho_p$ (yellow). Snapshots are given at four different times during the pulse interaction with a molecular system.}
    \label{fig:rhop_ilustr}
\end{figure}

Eq.~\eqref{eq:dens_prop3b} provides a clear interpretation of the excitation process: for every time $t^\prime$ of the laser pulse, the initial ground-state nuclear density is first multiplied by the transition dipole moment and the Wigner representation of the pulse and then promoted to the excited electronic state to be propagated from time $t^\prime$ to a desired time $t$. The propagation is governed by the time-independent Hamiltonian for the excited electronic state, without the explicit laser pulse. Integration over all times $t^\prime$ during the pulse until the current time $t$ reconstructs the full excited-state nuclear density at time $t$, while the ground-state density remains unperturbed. This whole process is further illustrated in Fig.~\ref{fig:rhop_ilustr}.

Hence, Eq.~\eqref{eq:dens_prop3b} offers a strategy to decouple the excitation process from the subsequent excited-state dynamics. By considering that the early times of the excited-state dynamics behave adiabatically, we can combine the description of photoexcitation discussed above to existing methods for excited-state molecular dynamics like fewest-switches surface hopping or \textit{ab initio} multiple spawning. The only required additional step consists of sampling the promoted nuclear density $\rho^\mathrm{cl}_p(\mvec{R},\mvec{P},t^\prime)$, but this is a simple task once we have the ground-state nuclear density, as we should soon see. At the end of our derivation, we would also like to state that the range of validity of Eq.~(\ref{eq:dens_prop3b}) is limited and a detailed discussion about its assumptions and approximations is provided in the SI. 

Before turning to the practical implementation of PDA, we briefly return as promised to the Wigner pulse representation -- the key component of Eq.~\eqref{eq:dens_prop3b}. Omitting their polarization, laser pulses can be either represented in the time domain as a time-dependent electric field $E(t)$ or in the frequency domain as a pulse spectrum $\Tilde{E}(\omega)$. These two representations are connected via a Fourier transform $\Tilde{E}(\omega) = \int_{-\infty}^\infty E(t)\mathrm{e}^{-i\omega t} \dd t$. 
The Wigner pulse representation $\mathcal{W}_E$ provides a way to obtain simultaneous temporal and frequency information by representing the pulse in both the time and frequency domain. As its cousin from quantum mechanics, the Wigner pulse representation does not have a proper physical meaning as such, but its integrated forms provide important insights. In particular, integration over frequency
\begin{align}
    &\int_{-\infty}^{\infty} \mathcal{W}_E(t,\omega) \dd \omega = 2\pi \left| E(t) \right|^2  \approx I(t)  \label{eq:wigint1}
\end{align}
yields the pulse intensity profile $I(t)$\footnote{For pulses in form $\varepsilon(t)\cos(\omega_0 t)$ the intensity is defined as $I(t) = \frac{1}{2}\epsilon_0 c n \varepsilon^2(t)$ where $\epsilon_0$ is the permittivity, $c$ is the speed of light and $n$ is the material refractive index.\cite{DIELS20061}} while integration over time
\begin{align}
    &\int_{-\infty}^{\infty} \mathcal{W}_E(t,\omega) \dd t = \left| \Tilde{E}(\omega) \right|^2 \approx S(\omega) \label{eq:wigspec1}
\end{align}
gives the spectral intensity $S(\omega)$.\cite{DIELS20061} Both $I(t)$ and $S(\omega)$ are the quantities actually accessible experimentally (unlike $E(t)$ and $\Tilde{E}(\omega)$), providing a more direct connection to laser experiments. Similarly to the Wigner representation of density, $\mathcal{W}_E$  can also acquire negative values making it rather quasiprobability than probability function. The occurrence of negative values hinders practical sampling from the distribution, yet we will discuss later that $\mathcal{W}_E$ is positive for the usual Gaussian envelope and only minor problems appear for other standard pulse envelopes. 

Let us now illustrate some of the quantities discussed above with the practical example of sodium iodide (NaI) interacting with a Gaussian laser pulse. 
The parameters for the sodium iodide Hamiltonian are provided in the SI. We define the electric field of the Gaussian laser pulse as\footnote{We note that defining pulses in the form of an envelope multiplied by an oscillating phase is generally not in line with Maxwell equations and works only for pulses with $\tau>2$~fs in the visible region and $\tau>1.2$~fs for UV pulses.
More details about the limit of the pulse envelope formulation and its derivation are provided in the SI.}
\begin{equation}
    E(t) = \exp\left( -2 \ln2 \frac{t^2}{\tau^2}\right) \cos\left(\omega_0 t\right) \, ,
\end{equation}
where $\omega_0$ is the frequency of oscillations and $\tau$ corresponds to the full width at half maximum (FWHM) parameter of the pulse intensity $I(t) \approx \varepsilon^2(t) = \exp\left( -4 \ln2 \frac{t^2}{\tau^2}\right)$. We advocate for using the FWHM parameter for the intensity profile rather than for the field envelope $\varepsilon$ (i) because the intensity is the experimentally accessible and reported quantity and (ii) because the population transfer to the excited electronic state is proportional to the intensity. Such a definition of $\tau$ hence minimizes disparities between experiments and theoretical investigations. 

\begin{figure}[ht!]
    \centering
    \includegraphics[width=0.68\textwidth]{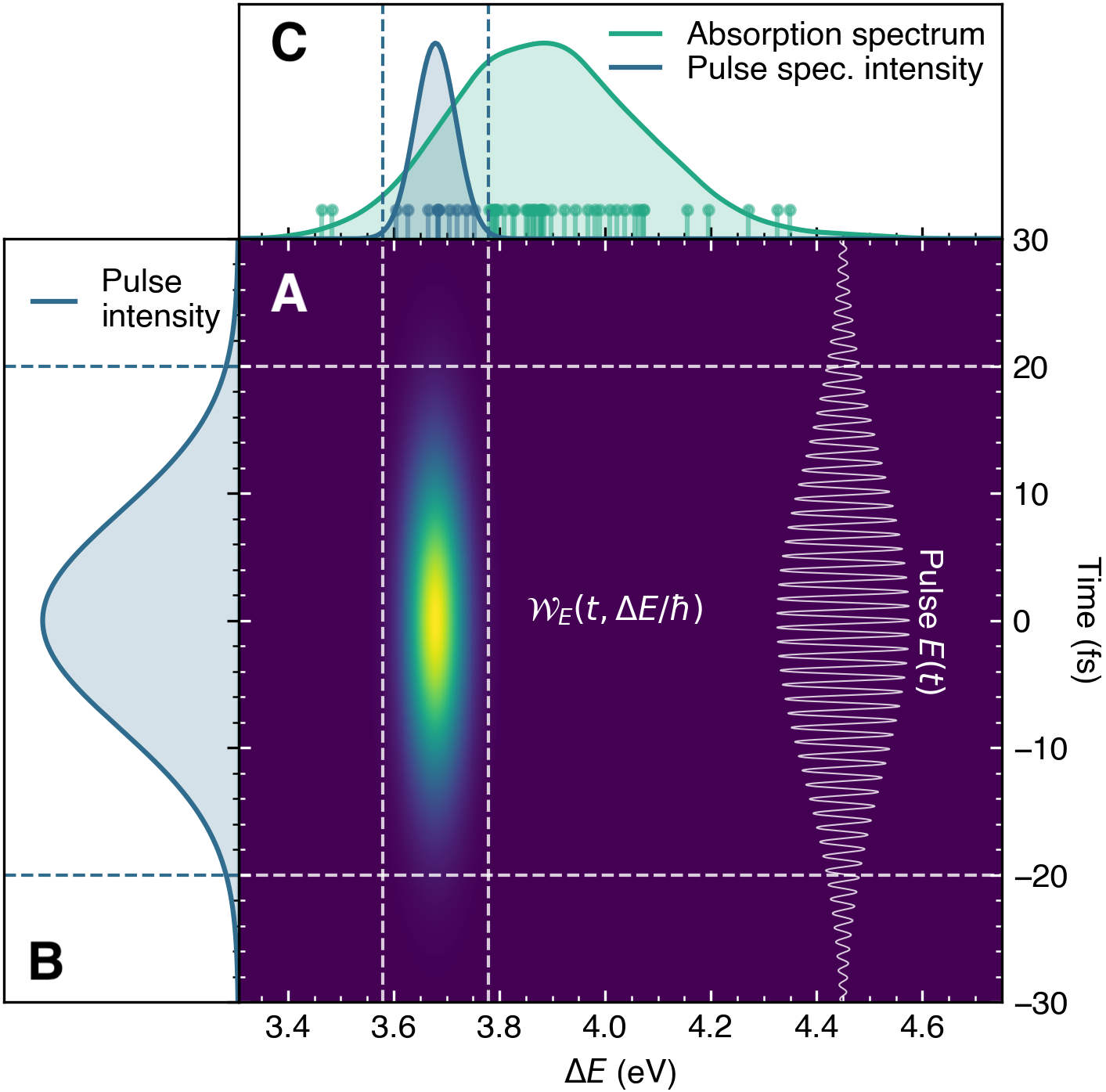}
    \caption{Illustration of the Wigner pulse representation and associated quantities for a laser pulse applied to sodium iodide. (A) Wigner pulse representation $\mathcal{W}_E$ for a Gaussian pulse with $\tau=20\,\text{fs}$ and $\hbar\omega_0 = 3.6792\,\text{eV}$. Dashed lines correspond to a region containing twice the FWHM parameters of pulse intensity and spectral intensity. The pulse electric field is plotted as an inset. (B) Projection of $\mathcal{W}_E$ on the time axis (pulse intensity $I$). (C) Projection of $\mathcal{W}_E$ on the frequency axis (pulse spectral intensity $S$) and the absorption spectrum of NaI. The sticks represent a sample of 50 ground-state geometries sampled from the ground-state Wigner distribution, with the blue colour used to represent geometries in resonance with the pulse.}
    \label{fig:pwig_ilustr}
\end{figure}

The Wigner pulse representation $\mathcal{W}_E$ for a specific pulse with $\tau=20\,\text{fs}$ and $\hbar\omega_0 = 3.68\,\text{eV}$ (corresponding to $\Delta E^\mathrm{el}_{eg}$ at a NaI distance of 2.74 \AA) is represented in Figure~\ref{fig:pwig_ilustr}A alongside the pulse intensity $I(t)$ (panel B), spectral intensity $S(\omega)$ (panel C), and oscillating electric field $E(t)$ (inset). The plotted $\mathcal{W}_E$ depicts the probability for a molecular geometry with excitation energy $\Delta E$ to be promoted to the excited state throughout the pulse duration. The spectral intensity for a 20-fs pulse is already quite narrow, targeting only a part of the NaI absorption spectrum. Thus, only a fraction of the ground-state probability density can potentially be excited. 

Having now a mental picture of the Wigner pulse representation and an equation for the promoted nuclear density (Eq.~\eqref{eq:dens_prop3b}), we can devise PDA -- a practical implementation for sampling initial conditions for nonadiabatic dynamics that incorporate implicitly the effect of a laser pulse, in other words, sampling the promoted nuclear density $\rho^\mathrm{cl}_p$. We start by the usual sampling of the ground-state nuclear probability density $\rho^\mathrm{cl}_g(\mvec{R},\mvec{P})$, producing a set of $N_g$ pairs of nuclear positions and momenta $\{\mvec{R}_i,\mvec{P}_i\}_{i=1}^{N_g}$. Different techniques can be used for this task, e.g., the harmonic Wigner sampling\cite{Wigner1932,Hillery1984} (available in most codes for nonadiabatic molecular dynamics) or \textit{ab initio} molecular dynamics with quantum thermostat.\cite{Ceriotti2009,Ceriotti2010,prlj2023deciphering} Then, excitation energies ($\Delta E_{eg}(\mvec{R})$) and transition dipole moments ($\Vec{\mu}_{eg}(\mvec{R})$) to the excited electronic state of interest must be obtained for each sampled nuclear configuration $\mvec{R}_i$ -- this process is also commonly performed to calculate a photoabsorption cross-section from the sampled geometries using the nuclear ensemble approach.\cite{Crespo-Otero2012} Having collected this information, we now take the first step beyond the standard workflow and randomly select an excitation time $t^\prime$ (from a time window surrounding the laser pulse of interest) and a position-momentum pair $\{\mvec{R}_i,\mvec{P}_i\}$. Using this random excitation time $t^\prime$ and $\{\mvec{R}_i,\mvec{P}_i\}$, a transition probability $p=|\Vec{\mu}_{eg}(\mvec{R}_i)\cdot\Vec{E}_0|^2\mathcal{W}_E(t^\prime, \Delta E_{eg}(\mvec{R}_i))$  can be obtained based on Eq.~\eqref{eq:finalden}. This probability is compared to a uniformly generated random number and, if this random number is smaller than the transition probability, the initial condition $\{\mvec{R}_i,\mvec{P}_i,t^\prime\}$ is accepted. This process of random selection is iterated until a desired number of initial conditions $N_p$ is created $\{\mvec{R}_j,\mvec{P}_j,t^\prime_j\}_{j=1}^{N_p}$ (see Algorithm~S1 in the SI for an algorithmic representation of the process).\footnote{We use index $j$ for the initial conditions generated by PDA, while the index $i$ is reserved for the ground-state sampling of position-momentum pairs $\{\mvec{R}_i,\mvec{P}_i\}$.} 
The excitation time $t^\prime_j$ stands for the actual time within the laser pulse envelope when the trajectory is initiated in the excited electronic state with \{$\mvec{R}_j$,~$\mvec{P}_j$\} (in contrast, the vertical sudden approximation would initiate \textit{all} trajectories at the excitation time $t'=0$). 
We note that some \{$\mvec{R}_i$,~$\mvec{P}_i$\} pairs may not be selected at all for excitation, while some other \{$\mvec{R}_i$,~$\mvec{P}_i$\} pairs may get excited at various times during the pulse; in other words, we can assign no or multiple excitation times $t^\prime$ to any given \{$\mvec{R}_i$,~$\mvec{P}_i$\} pair. The algorithm is straightforwardly extended to photoexcitation toward multiple excited electronic states by randomly selecting also the excited state (see SI). A user-friendly Python implementation of the algorithm is available\cite{pypipromdens, pda_code}, and a detailed description of the code is provided in the SI.

Once the initial conditions $\{\mvec{R}_j,\mvec{P}_j,t^\prime_j\}_{j=1}^{N_p}$ are sampled, the trajectory-based nonadiabatic dynamics simulation can be initiated in the electronic state of interest from unique position-momentum pairs \{$\mvec{R}_j$,~$\mvec{P}_j$\} within the initial conditions, starting at time 0. 
Finally, these resulting simulations should be shifted to their respective excitation times $t^\prime_j$ before analysis. 
We propose to start the simulations at time 0 and then shift them to $t^\prime_j$ to avoid repeated calculations of the same simulation just shifted to a different initial time. This way, the unique trajectories are used multiple times by shifting them to different $t^\prime_j$s. 
Note that the trajectory itself is considered fixed in the ground state until time $t^\prime_j$, when it is promoted to the excited electronic state. 

We now move to a numerical demonstration of PDA performance. To do so, we propose to compare the results of numerically-exact quantum dynamics (QD) simulations of NaI including explicitly a laser pulse to those obtained with FSSH using PDA, that is, incorporating the laser pulse implicitly in the initial conditions as described above. 

\begin{figure}[ht!]
    \centering
    \includegraphics[width=0.95\textwidth]{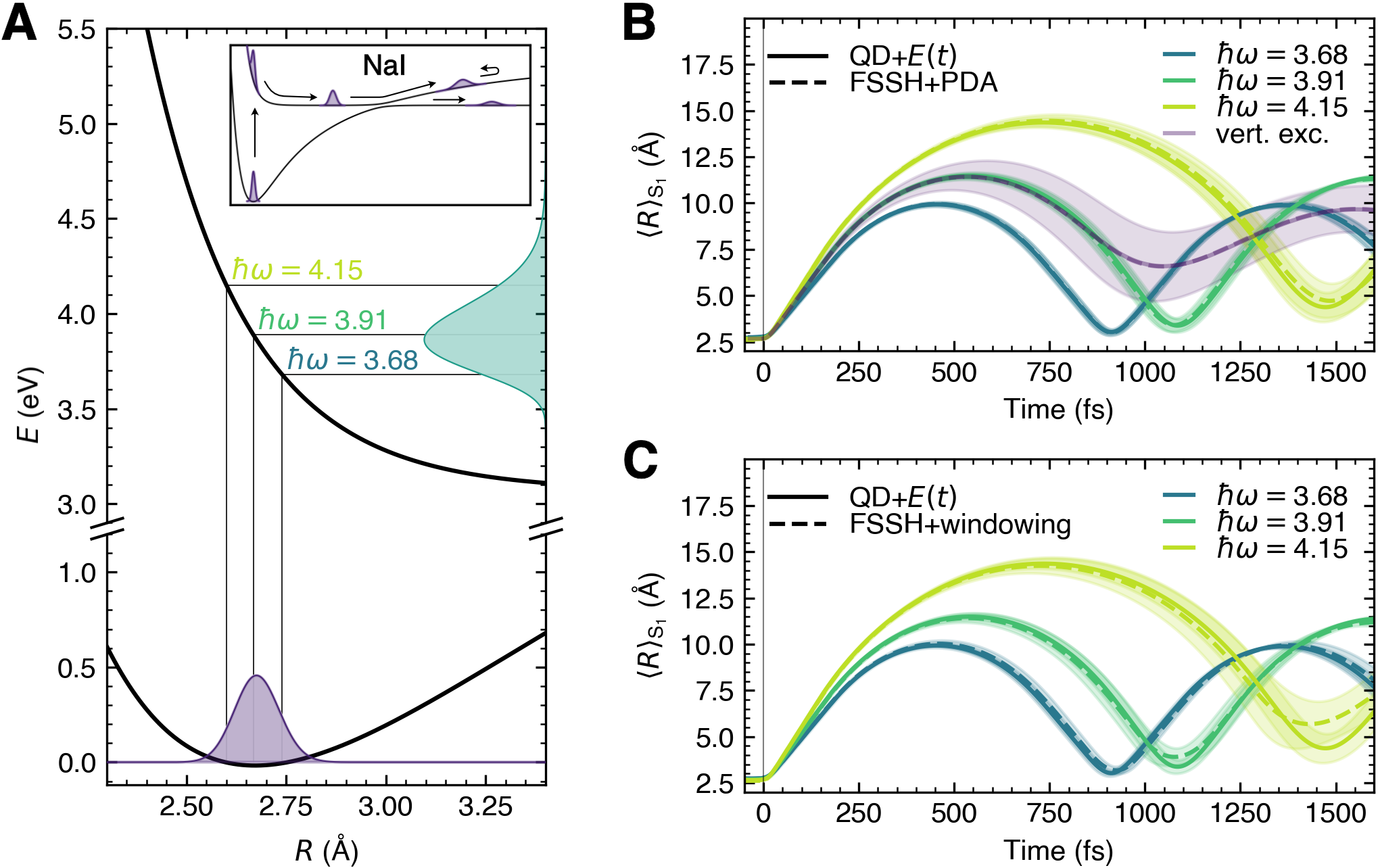}
    \caption{Photoexcitation of NaI with laser pulses of different frequencies, comparing QD with an explicit laser pulse and FSSH with PDA. (A) Potential energy curves of NaI (full range with sketched nuclear wavepacket evolution is given in the inset) alongside the ground-state density (violet) and the absorption spectrum (green) calculated with the nuclear ensemble method. The different frequencies $\hbar\omega$ of the applied laser pulse are depicted. (B) Expectation values of the NaI bond length in $S_1$ for three different laser pulse frequencies, comparing quantum dynamics with an explicit 20-fs laser pulse (solid lines) and PDA combined with FSSH nonadiabatic dynamics (dashed lines). The shaded area represents $\langle \Delta R \rangle_{\mathrm{S}_1} = \sqrt{\langle R^2 \rangle_{\mathrm{S}_1} - \langle R \rangle_{\mathrm{S}_1}^2}$ of the nuclear wavepacket (QD) or trajectories (FSSH). Simulations considering vertical excitation are provided as validation. (C) Same as in panel B but for FSSH using a simple windowing approach combined with a time convolution (dashed lines).}
    \label{fig:nai_omega}
\end{figure}

We start by comparing QD with an explicit laser pulse and FSSH+PDA for the photoexcitation of NaI resulting from three 20-fs laser pulses with different frequencies (Figure~\ref{fig:nai_omega}A). Our comparison is based on the expectation value of the NaI bond length in the first adiabatic excited state $S_1$, $\langle R\rangle_{S_1}$, and its standard deviation, $\langle \Delta R \rangle_{\mathrm{S}_1}$.
FSSH combined with PDA is in excellent agreement with the result obtained by QD with the explicit laser pulse (Figure~\ref{fig:nai_omega}B, note the perfect overlap between dashed and solid lines). Increasing the pulse frequency generates an excited nuclear wavepacket that takes longer to return to the Franck-Condon region following photoexcitation. The 20-fs pulses promote only small portions of the ground-state nuclear density, as seen in Figure~\ref{fig:pwig_ilustr} depicting the low-frequency pulse. The resulting excited nuclear wavepacket is promoted on different regions of the potential energy curve depending on the laser pulse frequency, each possessing then a different total energy resulting in different periods of oscillations. The violet dashed and solid lines in Figure~\ref{fig:nai_omega}B represent the dynamics obtained from a vertical (or sudden) excitation\footnote{Vertical (or sudden) excitation assumes that the whole unchanged ground-state nuclear density is promoted to the excited electronic state instantaneously at time $t$, i.e., $\rho_p = |\Vec{\mu}_{eg}\cdot\Vec{E}_0|^2\delta(t) \rho_g$.} of the ground-state nuclear wavefunction (both in QD and FSSH). While this set of simulations serves as a test of consistency between QD and FSSH (nearly indistinguishable results are obtained), they also clearly spotlight the remarkable differences between the dynamics triggered by the vertical (sudden) excitation and those obtained by considering the effect of a laser pulse. These results make it clear that using different laser pulses triggers excited-state dynamics with distinct timescales and incorporating such effects is crucial for a proper comparison with an experiment: here, the range of oscillation periods varies from 900 to 1500~fs depending on the laser pulse. 

When nonadiabatic dynamics simulations account for a laser pulse implicitly, they usually do so in a simple way by imposing an energy window on the selection of initial conditions (symbolized by the vertical dashed lines in Figure~\ref{fig:pwig_ilustr}). Testing this windowing approach reveals that, although less accurate than PDA, it still captures the major effects created by the laser pulse and brings a large improvement over the vertical excitation approach (Figure~\ref{fig:nai_omega}C). The definition of this energy window is, nevertheless, arbitrary. PDA allows us to validate a more rigorous scheme to perform a simple windowing. Considering the properties of the Wigner pulse representation given in Eqs.~\eqref{eq:wigint1} and~\eqref{eq:wigspec1}, we can approximate it as
\begin{equation}
    \mathcal{W}_E(t,\omega) \approx I(t) S(\omega)
    \label{eq:wig_approx}
\end{equation}
and recast Eq.~\eqref{eq:dens_prop3} into the following, simplified form
\begin{equation}
    \rho^\mathrm{cl}_e(\mvec{R},\mvec{P},t) = \frac{1}{\hbar^2} \int_{-\infty}^{\infty} I(t-t^{\prime})\mathrm{e}^{\mathcal{L}^\mathrm{cl}_e t^{\prime}} \left[ |\Vec{\mu}_{eg}(\mvec{R})\cdot\Vec{E}_0|^2 S(\Delta E^\mathrm{el}_{eg}(\mvec{R})/\hbar)  \rho^\mathrm{cl}_g(\mvec{R},\mvec{P}) \right] \dd t^{\prime} \, ,
    \label{eq:dens_prop4}
\end{equation}
where the integral has a structure of convolution. Eq.~\eqref{eq:dens_prop4} provides a clear recipe for a simple windowing and convolution approach that we coined PDAW (promoted density approach for windowing). We can assign a weight 
\begin{equation}
    w_i=|\Vec{\mu}_{eg}(\mvec{R}_i)\cdot\Vec{E}_0|^2 S(\Delta E^\mathrm{el}_{eg}(\mvec{R}_i)/\hbar)
\end{equation} 
to each position-momentum pair $\{\mvec{R}_i,\mvec{P}_i\}$ from the ground-state sampling. Then, we propagate $\{\mvec{R}_i,\mvec{P}_i\}$ in the excited state and calculate an observable $\mathcal{O}_i(t)$ for each trajectory. 
The total time-dependent observable $\mathcal{O}$ is then calculated by summing the weighted (and normalized) observables obtained from each initial condition and convoluting the result with the adequate form of the normalized pulse intensity $\Bar{I}(t)=I(t)/\int_{-\infty}^{\infty}I(t^{\prime})\dd t^{\prime}$: 
\begin{equation}
    \mathcal{O}(t) = \int_{-\infty}^{\infty} \Bar{I}(t-t^{\prime}) \frac{\sum_i w_i \mathcal{O}_i(t^\prime)}{\sum_i w_i} \dd t^{\prime} \, .
\end{equation}
We emphasize that the weights $w_i$ are based on the pulse spectral intensity S($\omega$) and not the pulse spectrum $\Tilde{E}(\omega)$. Applying the PDAW scheme to FSSH for the three Gaussian laser pulses described in Fig.~\ref{fig:nai_omega} outperforms the standard windowing and yields the same results as FSSH combined with PDA, except for dynamics during the pulse as PDAW is based on Eq.~\eqref{eq:dens_prop3} and not Eq.~\eqref{eq:dens_prop3b} (see SI). However, we note that PDAW is still an approximation to PDA and Eq.~\eqref{eq:wig_approx} will not be valid in general, e.g. for chirped pulses with long duration (see SI).

\begin{figure}[ht!]
    \centering
    \includegraphics[width=0.7\textwidth]{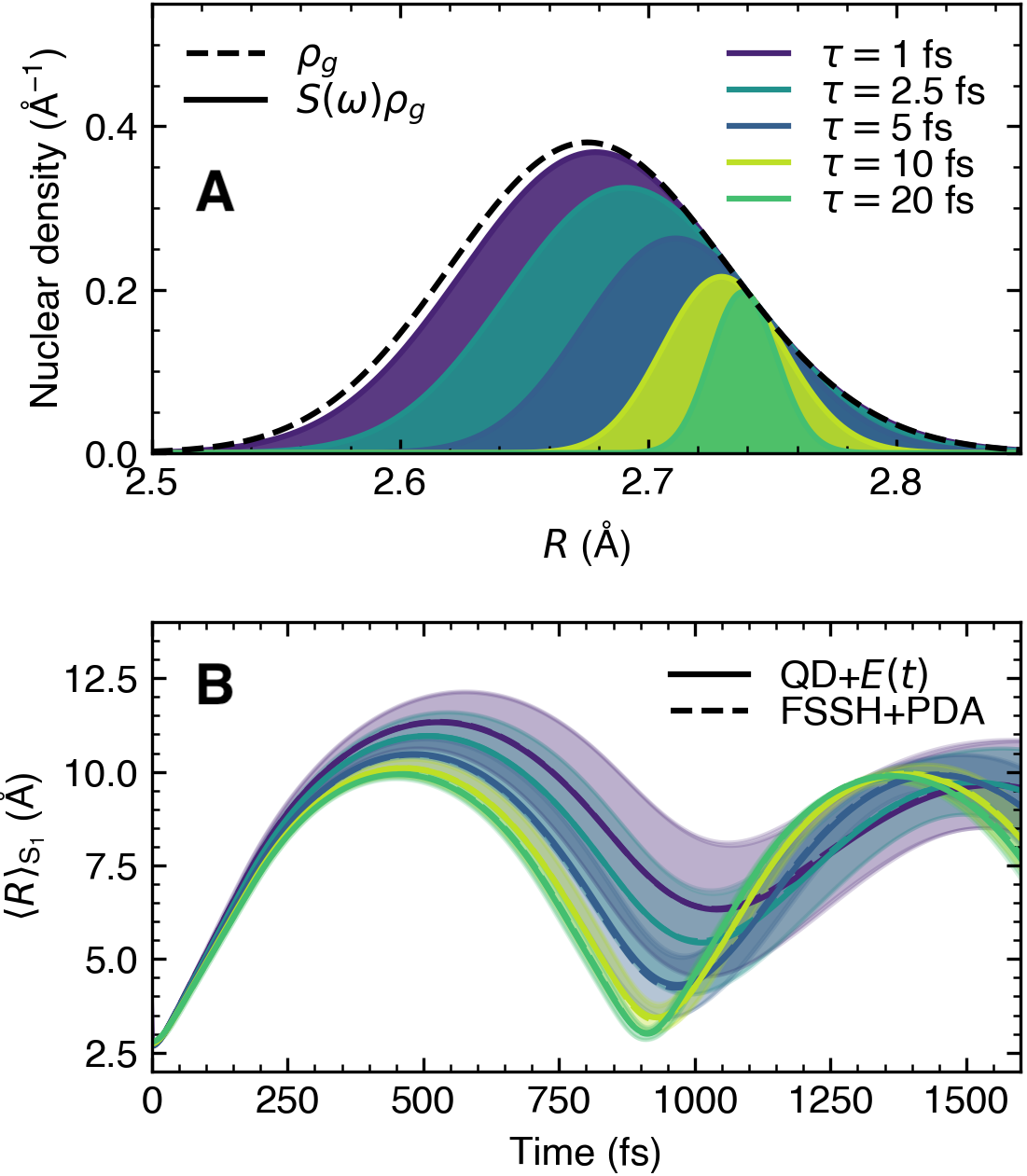}
    \caption{Photoexcitation of NaI with laser pulses of different durations, comparing QD with an explicit laser pulse and FSSH with PDA. (A) The ground-state nuclear density $\rho_g$ (dashed line) and its product with the spectral intensity $S(\omega)$ (filled areas) for pulses of different duration $\tau$. (B): Expectation values of the NaI bond length in $S_1$ for the different laser pulse durations given in (A), all with $\hbar\omega_0 = 3.6792\,\text{eV}$, for QD with an explicit laser pulse (solid lines) and FSSH using PDA (dashed lines). The shaded area represents $\langle \Delta R \rangle_{\mathrm{S}_1}$ of the nuclear wavepacket (QD) or trajectories (FSSH). We note that the dashed lines are hardly visible as they almost perfectly overlap the corresponding solid lines.}
    \label{fig:nai_fwhm}
\end{figure}

How good is PDA to describe pulses of different durations? Let us test the approach for a series of Gaussian pulses with parameters $\tau=1$, 2.5, 5, 10, and 20~fs considering the central frequency $\hbar\omega_0 = 3.6792\,\text{eV}$ (Figure~\ref{fig:nai_fwhm}). In Figure~\ref{fig:nai_fwhm}A, we represent the product of the pulse spectral intensity $S$ with the ground-state density $\rho_g$, demonstrating the portions of $\rho_g$ targeted with the laser pulses of different durations. While the shortest 1-fs laser pulse\footnote{A 1-fs laser pulse with a central frequency corresponding to $\hbar\omega_0 = 3.6792\,\text{eV}$ is just at the boundary of being valid within a pulse envelope formalism (see SI).} encompasses almost the whole ground-state density, the longest 20-fs pulse strikes only a small part of it. Again, PDA combined with FSSH leads to results in excellent agreement with QD using an explicit laser pulse for all pulses tested (Figure~\ref{fig:nai_fwhm}B), despite the strong effect of the laser pulse on the resulting excited-state dynamics.  While the 1-fs pulse leads to a nonadiabatic dynamics closely matching the one obtained after a vertical excitation (depicted in Figure~\ref{fig:nai_omega}B), the 20-fs pulse creates a more confined excited-state nuclear wavepacket (compare shaded areas in Figure~\ref{fig:nai_fwhm}B). These results once again strongly advocate for including the effect of a laser pulse in nonadiabatic molecular dynamics.

So far, we have shown that PDA combined with FSSH leads to nonadiabatic dynamics simulation in excellent agreement with quantum dynamics including an explicit laser pulse. Yet we note that the selected cases are all within PDA's approximations, such as short pulse duration or positive pulse Wigner representation,  summarized in SI. To stress its applicability, we pushed the approach out of its comfort zone in different ways. We briefly summarize here our findings and the interested reader can refer to the SI for the results and comparisons. PDA combined with FSSH provides excellent results even for 100-fs long laser pulse, but disparities with the QD results start to appear for a 500-fs pulse with FSSH. These disparities are not \textit{per se} due to a limitation of PDA for describing such long pulses but are caused by the appearance of quantum-interference effect in the dynamics when the excited-state wavepacket returns to the Franck-Condon region while the interaction with the laser pulse is still going on. We also tested different envelopes for the laser pulse, such as the Lorentzian envelope which exhibits negative regions in its Wigner pulse representation. Both strategies introduced for dealing with negative values -- either ignoring them or taking their absolute value -- perform well although the agreement is not as perfect as for the Gaussian pulses. Notably, our test simulations show that PDA can also capture excited-state dynamics during the pulse, for which the running equations are in principle not derived. PDAW proved to be also applicable for excited-state dynamics involving Lorentzian pulses, even if it lacks the dynamical effects observed during the pulse with PDA. Finally, we confirmed the applicability of the PDA strategy for chirped laser pulses. 

\begin{figure}[ht!]
    \centering
    \includegraphics[width=1\textwidth]{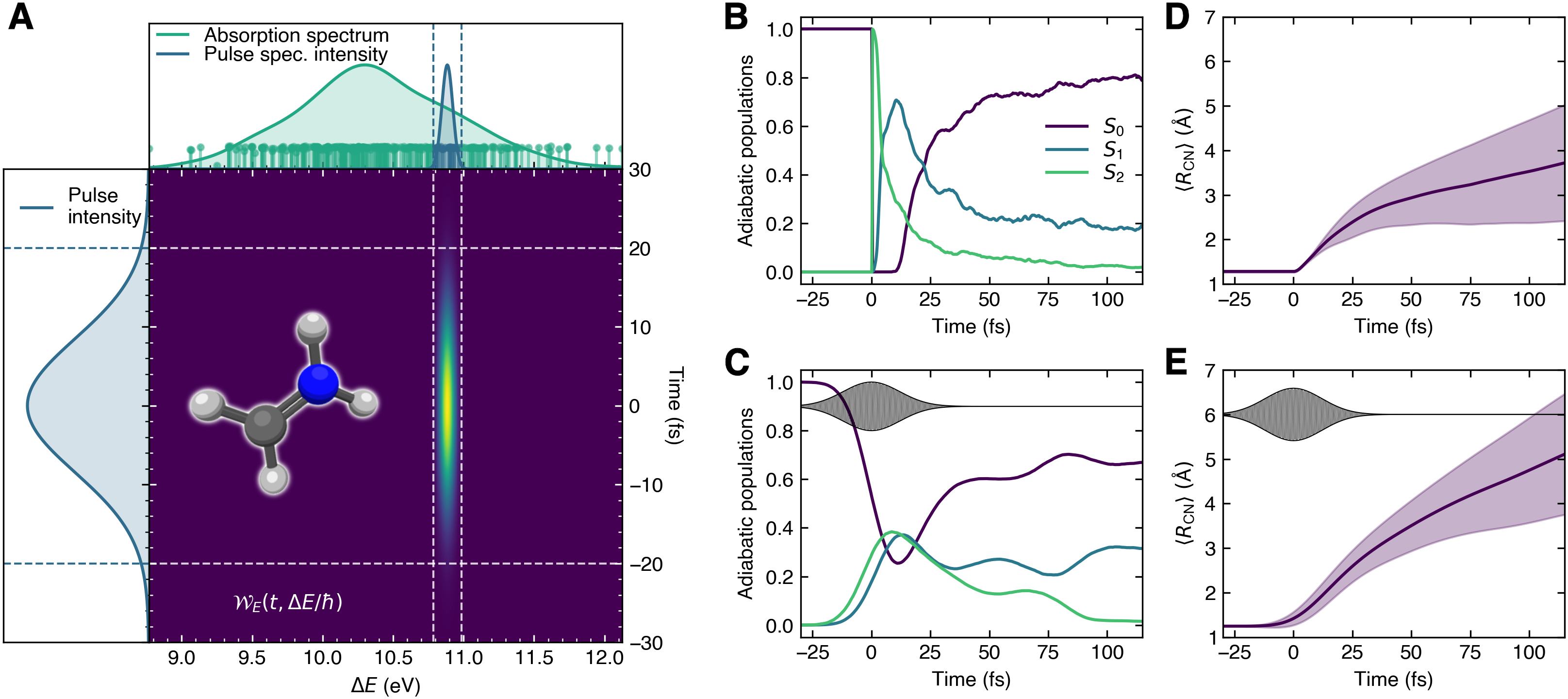}
    \caption{Photodynamics of protonated formaldimine, comparison between FSSH and the vertical (sudden) approximation and FSSH using PDA.(A) The Wigner pulse representation with pulse intensity, spectral intensity, and absorption spectrum of protonated formaldimine. The excitation energies of the 500 nuclear position-momentum pairs are represented as sticks with a height proportional to $|\mu_{S_2S_0}(\mvec{R}_i)|^2$. Blue sticks are used to depict the pairs that were selected by PDA while green sticks indicate pairs that were not promoted by the pulse. (B) Time trace of the adiabatic electronic populations for FSSH with a vertical excitation. (C) Time trace of the adiabatic electronic populations for FSSH with PDA. (D) Mean carbon-nitrogen bond length ($R_\text{CN}$) during the FSSH dynamics initiated from a vertical excitation. (E) Mean carbon-nitrogen bond length ($R_\text{CN}$) during the FSSH dynamics initiated from PDA.}
    \label{fig:formald}
\end{figure}

As a final example, we wish to illustrate the applicability of PDA to a larger molecular system, combined with on-the-fly FSSH dynamics. We selected protonated formaldimine as an example -- a molecule well-known to the community and often used for benchmarking nonadiabatic dynamics\cite{Suchan2020,Curchod2020} -- and simulated its interaction with a Gaussian laser pulse ($\omega_0=0.40$~a.u. and $\tau=20$~fs). The photodynamics of protonated formaldimine is particularly interesting as the population transfer from the photoexcited $S_2$ states is notoriously fast. 
To demonstrate that PDA can also be used as a post-processing tool for FSSH dynamics, we reuse the 500 position-momentum pairs $\{\mvec{R}_i,\mvec{P}_i\}_{i=1}^{500}$ and corresponding FSSH trajectories from our previous work considering vertical excitation.\cite{Suchan2020} From these position-momentum pairs,  we generated a set of 50000 initial conditions $\{\mvec{R}_j,\mvec{P}_j,t^\prime_j\}_{j=1}^{50000}$ using PDA that contains only 63 unique positions and momenta, i.e. only 63 FSSH trajectories are required (see Figure~\ref{fig:formald}). These 63 trajectories were shifted to different initial times $t^\prime_j$ creating a new set of 50000 trajectories used for the analysis. 

The photodynamics of protonated formaldimine is significantly altered by including an implicit laser pulse within PDA. The FSSH simulations with a vertical excitation provide a lifetime for the $S_2$ electronic state that is comparable to the pulse duration used with PDA. As a result, the population of the $S_2$ electronic state only reaches a maximum of only 0.4 when the 20-fs laser pulse is included, as the sink of the $S_2$ population (via nonadiabatic processes) is faster than the source of population from $S_0$ caused by the laser pulse. The overall population dynamics is also more spread in time within PDA, exemplifying the importance of pulse duration effects on ultrafast processes like internal conversion. The time evolution of the mean bond length ($R_\text{CN}$) is also affected by the finite-energy spectrum of the 20-fs laser pulse, leading to a faster extension in comparison to the FSSH dynamics invoking a vertical excitation. While none of these effects are surprising, they clearly demonstrate the profound consequences of adequately including the effect of a laser pulse in nonadiabatic molecular dynamics. Furthermore, the results highlight that PDA can be used to infer the laser pulse effects in a post-processing manner from already calculated simulations considering vertical excitation. This can be particularly useful when studying novel molecules for which experiments have yet to be conducted.

In conclusion, we proposed a formalism coined PDA for the implicit inclusion of arbitrary laser pulses in nonadiabatic molecular dynamics at the level of the initial conditions. The core component of this method is the Wigner pulse representation used to calculate an (approximate) promoted nuclear density for the subsequent excited-state dynamics. Sampling the promoted density is the only additional step required in comparison to the standard approach invoking a vertical excitation. We have demonstrated the performance and possible limitations of the formalism by simulating the photoexcitation of NaI with FSSH and PDA using a broad range of laser pulses, leading to excellent agreements with QD simulations including explicitly the laser pulse. The excellent performance of FSSH+PDA contrasts with the poor performance of FSSH when coupled to an explicit laser pulse, as reported in earlier works.\cite{bajo2014interplay,Mignolet2019b} We also derived the approximate PDAW formalism that sets the simpler energy windowing and intensity convolution strategy on firm ground. The applicability of the PDA approach to molecular systems (and its use as a postprocessing tool, that is, allowing to test the effect of various laser pulses using the same set of nonadiabatic trajectories) was demonstrated with the photoexcitation of protonated formaldimine. Both PDA and PDAW are implemented in a user-friendly Python code \verb|promdens.py| available as a Python package in the PyPI repository\cite{pypipromdens} or on GitHub\cite{pda_code} (a full description of the code and its availability is provided in the SI), using as an input only the results of a spectrum calculation obtained with the nuclear ensemble method. The technique is easy to extend to multiple electronic states, yet further developments are needed to describe a coherent superposition of electronic states within the present framework. Overall, the results of our simulations stress the importance of including the laser pulse in nonadiabatic molecular dynamics and the potential dangers of using the vertical (sudden) excitation. 

\begin{acknowledgement}

The authors thank Daniel Hollas for comments on the manuscript and assistance with the PDA code. This project has received funding from the European Research Council (ERC) under the European Union's Horizon 2020 research and innovation programme (Grant agreement No. 803718, project SINDAM) and the EPSRC Grants EP/V026690/1, EP/Y01930X/1, and EP/X026973/1. J. J. and P.S. thank the Czech Science Foundation for the support via grant number 23-07066S. This work was supported from the grant of Specific university research -- grant No. A2\_FCHI\_2024\_057. 

\end{acknowledgement}

\begin{suppinfo}

Derivation of the central equations of the promoted density approximation with definitions of all required quantities. 
Full computational details. Extended tests of the methodologies presented in this work (PDA vs PDAW vs windowing, chirped pulses, long laser pulses, Lorentzian envelopes). Description of the PDA code, its input, and its usage. (PDF file)

\end{suppinfo}

%\bibliography{references.bib}

\providecommand{\latin}[1]{#1}
\makeatletter
\providecommand{\doi}
  {\begingroup\let\do\@makeother\dospecials
  \catcode`\{=1 \catcode`\}=2 \doi@aux}
\providecommand{\doi@aux}[1]{\endgroup\texttt{#1}}
\makeatother
\providecommand*\mcitethebibliography{\thebibliography}
\csname @ifundefined\endcsname{endmcitethebibliography}
  {\let\endmcitethebibliography\endthebibliography}{}

\end{document}

\maketitle
\clearpage
\singlespacing
{ \hypersetup{hidelinks} \tableofcontents }

\clearpage
\section{Intensity, spectrum, and Wigner representation of a laser pulse}

In this Section, we aim to review some basic properties of laser pulses and justify facts mentioned in the main text. We start by defining the time-dependent electric field of the laser pulse as $\Vec{E}(t)=\Vec{E}_0 E(t)$, where $\Vec{E}_0$ is the electric field amplitude $E_0$ multiplied by the polarization vector $\Vec{\lambda}$ of the field. In the following, we will focus on the time-dependent complex scalar field $E(t)$,
\begin{equation}
    E(t) = \varepsilon(t)\mathrm{e}^{i\gamma(t)} \, ,
\label{eq:pulse}
\end{equation}
where $\varepsilon(t)$ is an envelope  and $\gamma(t)$ is a phase defined as
\begin{equation}
    \gamma(t) = \varphi_0 + \omega_0 t + \beta t^2 \, ,
\end{equation}
with $\varphi_0$ being the carrier-envelope phase, $\omega_0$ the carrier frequency, and $\beta$ the quadratic phase modulation or linear chirp parameter. The instantaneous frequency of a pulse, i.e., the frequency of the oscillations at every point in time, is then defined as the time derivative of the phase $\gamma$,
\begin{equation}
    \frac{\dd \gamma}{\dd t} = \Dot{\gamma} = \omega_0 + 2\beta t \, .
\label{eq:instfreq}
\end{equation}
For unchirped pulses ($\beta =0$), the instantaneous frequency equals to $\omega_0$.
We note that defining the field $E$ as a complex function has the advantage that the spectrum contains only the positive-frequency contributions while defining the field as a real function $E(t) = \varepsilon(t)\cos\gamma(t)$ creates a symmetric spectrum with unphysical negative-frequency components. So while we use the real electric field in the Hamiltonian for the quantum dynamics simulations, we rather work with the complex electric field when deriving its properties.

The intensity of the field, which is the experimentally measured quantity, is defined as 
\begin{equation}
    I(t) = \frac{1}{2}\epsilon_0 c n \varepsilon^2(t) \, ,
\label{eq:intensity}
\end{equation}
where $\epsilon_0$ is the permittivity, $c$ is the speed of light and $n$ is the material refractive index.\cite{DIELS20061}

The spectrum of the field comes from the Fourier transform of $E$,
\begin{equation}
    \Tilde{E}(\omega) = \mathcal{F}[E(t)] = \int_{-\infty}^\infty E(t)\mathrm{e}^{-i\omega t} \dd t = |\Tilde{E}(\omega)|\mathrm{e}^{i\phi(\omega)} \, ,
\end{equation}
where $|\Tilde{E}(\omega)|$ denotes the spectral amplitude and $\phi(\omega)$ spectral phase. However, the experimentally accessible quantity is rather the spectral intensity $S(\omega)$ than the spectrum. The spectral intensity can be derived as
\begin{equation}
    S(\omega) = \frac{\epsilon_0 c n}{\pi}\left|\Tilde{E}(\omega) \right|^2
\end{equation}
and equals to the square of the pulse spectrum $\Tilde{E}$ multiplied by constants.\cite{DIELS20061}

Let us now comment on the parameters used to determine and report pulses. The experimentally accessible quantities are the intensities $I(t)$ and $S(\omega)$. Therefore, the full width at half maximum (FWHM) values for the temporal resolution ($\tau$) and the spectral resolution ($\Omega$) are determined from the respective intensities. On the other hand, theoreticians are usually accustomed to working with the electric field $E(t)$ and pulse spectrum $\Tilde{E}(\omega)$ and report the FWHM quantities for them – this disparity often leads to a misunderstanding between theoreticians and experimentalists. The way one defines the FWHM parameters is arbitrary, but we need to remain consistent. In this work, we advocate the use of the FWHM parameters for intensities. Not only are they directly provided in experiments but intensities also determine the temporal convolution of theoretical quantities and the energy windowing, as we will show later. In practice, this means that when we set the parameter $\tau$, it is for the FWHM of the intensity $I(t)\approx\varepsilon^2(t)$ and not the field envelope $\varepsilon(t)$. The FWHM parameter of the envelope $\varepsilon(t)$ depends on the specific envelope form and is easy to convert.

In analogy to quantum mechanics, where we can represent the wavefunction in either position or momentum space but never in both simultaneously, we cannot directly represent the field in the time and frequency domains together due to their Fourier transform relationship. Nevertheless, we can borrow the concept of Wigner representation from quantum mechanics, which allows us to construct phase-space quantities, for laser pulses. The Wigner representation $\mathcal{W}_E$ of a laser pulse defined by $E(t)$ reads\cite{DIELS20061}
\begin{align}
    \mathcal{W}_E(t,\omega) &= \int_{-\infty}^{\infty} E\left(t+\frac{s}{2}\right) E^*\left(t-\frac{s}{2}\right) \mathrm{e}^{-i\omega s} \dd s \label{eq:def1} \\
    & = \frac{1}{2\pi}\int_{-\infty}^{\infty} \Tilde{E}\left(\omega+\frac{s}{2}\right) \Tilde{E}^*\left(\omega-\frac{s}{2}\right) \mathrm{e}^{i t s} \dd s \label{eq:def2}
\end{align}
and presents a simultaneous representation of the laser pulse in the time and frequency domain.

However, we need to approach the concept of a Wigner representation for laser pulses with the same care as in quantum mechanics. The Wigner representation of a laser pulse allows us to plot the correlation between the temporal and spectral domains, which is useful especially for chirped pulses, but one should not think about it as a physical object. For example, we cannot interpret the Wigner representation as a probability density of a certain frequency at a given time because $\mathcal{W}_E$ can acquire negative values. Measurable physical quantities can be obtained from integrated Wigner distribution: if we integrate $\mathcal{W}_E$ over the frequency domain,
\begin{align}
    \int_{-\infty}^{\infty} \mathcal{W}_E(t,\omega) \dd \omega &= \int_{-\infty}^{\infty} \int_{-\infty}^{\infty} E\left(t+\frac{s}{2}\right) E^*\left(t-\frac{s}{2}\right) \mathrm{e}^{-i\omega s} \dd s \dd \omega \\
    &=  \int_{-\infty}^{\infty} E\left(t+\frac{s}{2}\right) E^*\left(t-\frac{s}{2}\right) \int_{-\infty}^{\infty}\mathrm{e}^{-i\omega s} \dd \omega \dd s  \\
    &= 2\pi \int_{-\infty}^{\infty} E\left(t+\frac{s}{2}\right) E^*\left(t-\frac{s}{2}\right) \delta(s) \dd s  \\
    &= 2\pi \left| E(t) \right|^2 = 2\pi \varepsilon(t)^2 \approx I(t)\, , \label{eq:wigint1}
\end{align}
we get the square of the field envelope which is proportional to the intensity $I(t)$.
We used the identity $\int_{-\infty}^{\infty}\mathrm{e}^{-i\omega s} \dd \omega = 2\pi \delta(s)$. On the other hand, if we integrate over the time domain using the definition Eq.~(\ref{eq:def2}),
\begin{align}
    \int_{-\infty}^{\infty} \mathcal{W}_E(t,\omega) \dd t &= \frac{1}{2\pi} \int_{-\infty}^{\infty} \int_{-\infty}^{\infty} \Tilde{E}\left(\omega+\frac{s}{2}\right) \Tilde{E}^*\left(\omega-\frac{s}{2}\right) \mathrm{e}^{i t s} \dd s  \dd t  \\
    &= \frac{1}{2\pi}  \int_{-\infty}^{\infty} \Tilde{E}\left(\omega+\frac{s}{2}\right) \Tilde{E}^*\left(\omega-\frac{s}{2}\right) \int_{-\infty}^{\infty} \mathrm{e}^{i t s} \dd t  \dd s  \\    
    &= \int_{-\infty}^{\infty} \Tilde{E}\left(\omega+\frac{s}{2}\right) \Tilde{E}^*\left(\omega-\frac{s}{2}\right) \delta(s) \dd s  \\
    &= \left| \Tilde{E}(\omega) \right|^2 \approx S(\omega) \, , \label{eq:wigspec1}
\end{align}
we obtain the square of the pulse spectrum which is proportional to the spectral intensity $S(\omega)$. Similarly, various moments of the field, the instantaneous frequency, and other quantities can be obtained, for which we refer the reader to Ref.~\citenum{DIELS20061}.

The Wigner pulse representation $\mathcal{W}_E$ can be recast into a simpler form if the pulse is defined in terms of an envelope $\varepsilon$ and an oscillating phase $\gamma$ (see Eq.~(\ref{eq:pulse})). Substituting Eq.~(\ref{eq:pulse}) into Eq.~(\ref{eq:def1}) leads to
\begin{align}
    \mathcal{W}_E(t,\omega) &= \int_{-\infty}^{\infty} \varepsilon\left(t+\frac{s}{2}\right) \mathrm{e}^{i\gamma(t+s/2)}\varepsilon^*\left(t-\frac{s}{2}\right) \mathrm{e}^{-i\gamma(t-s/2)}\mathrm{e}^{-i\omega s} \dd s \\
    &= \int_{-\infty}^{\infty} \varepsilon\left(t+\frac{s}{2}\right) \varepsilon^*\left(t-\frac{s}{2}\right) \mathrm{e}^{-i[\omega s+\gamma(t-s/2) -\gamma(t+s/2)]} \dd s \\
    &= \int_{-\infty}^{\infty} \varepsilon\left(t+\frac{s}{2}\right) \varepsilon^*\left(t-\frac{s}{2}\right) \mathrm{e}^{-i[\omega -\Dot{\gamma}(t)]s} \dd s \label{eq:wigtowig} \, ,
\end{align}
where we employed a Taylor expansion to derive $\gamma(t-s/2) -\gamma(t+s/2) = -(\omega_0 + 2\beta t)s = -\Dot{\gamma}(t)s$. The time derivative of the phase $\Dot{\gamma}$ stands for the instantaneous frequency, see Eq.~(\ref{eq:instfreq}). We can now define a Wigner representation of the pulse envelope 
\begin{align}
    \mathcal{W}_\varepsilon(t,\omega) &= \int_{-\infty}^{\infty}  \varepsilon\left(t+\frac{s}{2}\right) \varepsilon^*\left(t-\frac{s}{2}\right) \mathrm{e}^{-i\omega s}  \dd s \, .
\end{align}
Note that the complex conjugation is not necessary for real envelopes yet we keep it to highlight the analogy to $\mathcal{W}_E$. Following from Eq.~(\ref{eq:wigtowig}), $\mathcal{W}_E$ and $\mathcal{W}_\varepsilon$ are connected via a simple relation:
\begin{equation}
    \mathcal{W}_E(t,\omega) =\mathcal{W}_\varepsilon(t,\omega-\Dot{\gamma}) \, .
\label{eq:wigtowig2}
\end{equation}
The concept of $\mathcal{W}_\varepsilon$ is less general than $\mathcal{W}_E$, as it works only for pulses defined as in Eq.~(\ref{eq:pulse}) and is not suitable for ultrashort pulses (see Section~\ref{sec:pulserep}). Yet, it is more efficient for numerical implementations and is utilized in our Python code (described in Section~\ref{sec:implement}). 

\clearpage
\section{Limits of the laser pulse definition via a pulse envelope}
\label{sec:pulserep}

Through this work, we represent laser pulses in the form of an envelope times an oscillating field, as in Eq.~(\ref{eq:pulse}). Although it is the most often used representation of a laser pulse, care is needed when using it. The following condition for a laser pulse in free space must hold
\begin{equation}
    \int_{-\infty}^\infty E(t) \dd t = 0 
\label{eq:cond1}
\end{equation}
to fulfill Maxwell equations. Several reasons for this condition to hold are comprehensively presented in Ref.~\citenum{Madsen2002}. We shall mention only one of them here: \textit{if the integral over the electric field is not equal to zero, a direct current is induced, which is unphysical in free space}. 

Whether a laser pulse is physical and fulfills the condition provided by Eq.~(\ref{eq:cond1}) can be easily verified from the zero frequency component of the pulse spectrum $\Tilde{E}$:
\begin{equation}
    \int_{-\infty}^\infty E(t) \dd t = \int_{-\infty}^\infty E(t) \mathrm{e}^{i(\omega=0)t }\dd t = \Tilde{E}(0) = 0 \, .
\label{eq:cond2}
\end{equation}
Thus, if a given definition of a laser pulse exhibits a nonzero spectrum at zero frequency, this laser pulse is not physical. As such, Eq.~(\ref{eq:cond2}) provides a quick check of whether a laser pulse definition is physical or not. 

\begin{figure}[b!]
    \centering
    \includegraphics[width=0.7\textwidth]{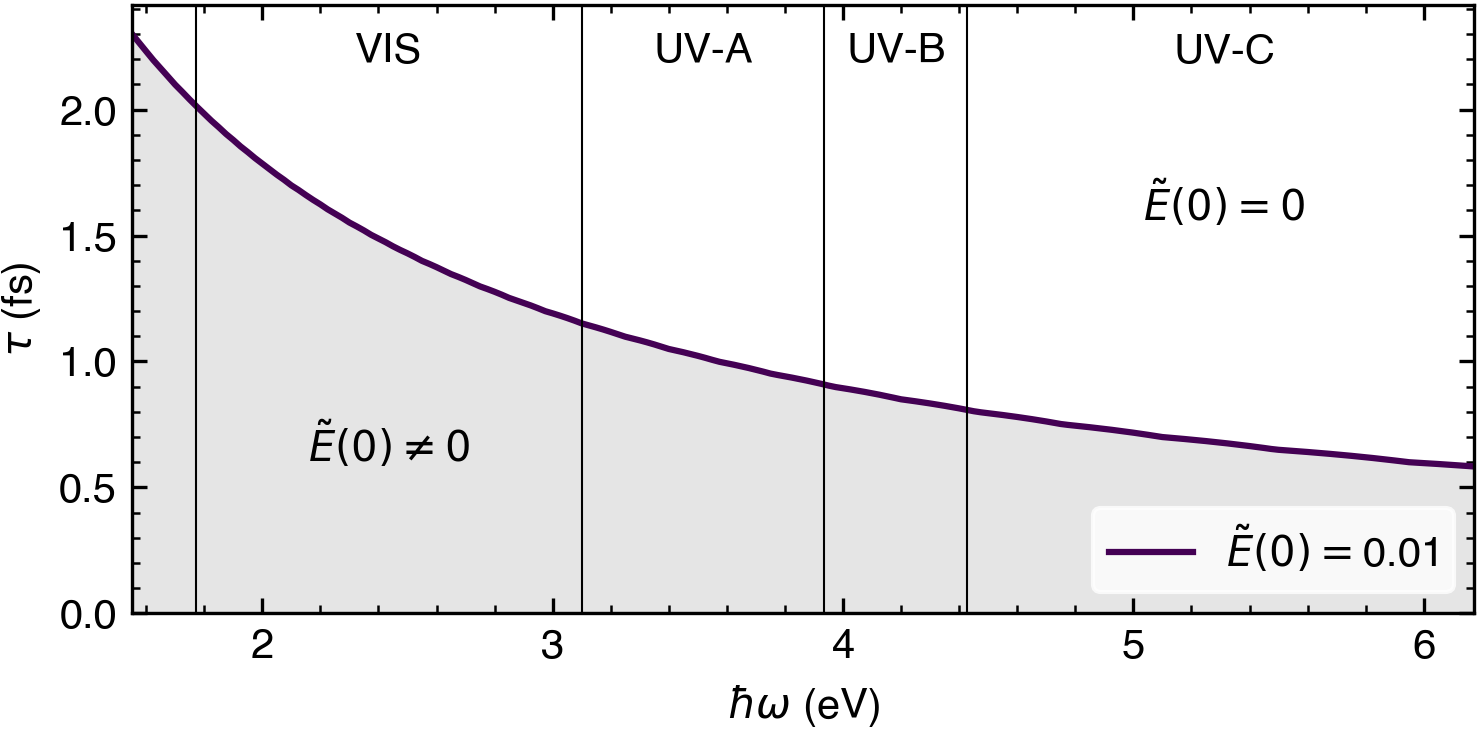}
    \caption{Validity of the Gaussian envelope representation for laser pulses. For each field frequency $\omega$, we calculated the envelope FWHM parameter ($\tau$) at which the normalized (maximum equal to 1) pulse spectrum has a value of 0.01 at zero frequency, i.e., $\Tilde{E}(0)=0.01$. If the value of $\Tilde{E}(0)$ is less than 0.01, we consider that the condition given by Eq.~(\ref{eq:cond1}) is fulfilled. If the value of $\Tilde{E}(0)$ is bigger than 0.01, we consider that the pulse is unphysical. The fit of the curve with an inverse function yields $\tau \mathrm{[fs]} = 3.5730048296 / \hbar\omega \mathrm{[eV]}$.}
    \label{fig:cond}
\end{figure}

While for longer pulses (more than a few femtoseconds), the definition of a laser pulse given by Eq.~(\ref{eq:pulse}) is valid based on the condition given in Eq.~(\ref{eq:cond1}), ultrashort pulses shorter than a femtoseconds approach the physical limits of such a representation. To further highlight this limitation, we have estimated the range of validity of laser pulses defined with a Gaussian envelope, $\varepsilon(t) = \exp\left( -2 \ln2 \frac{t^2}{\tau^2}\right)$ (see Figure~\ref{fig:cond}). Based on Figure~\ref{fig:cond}, we propose a rule of thumb for estimating whether a laser pulse can or cannot be represented as a Gaussian envelope time an oscillating field: \textit{If $\tau$ (in femtoseconds) for the Gaussian envelope is larger than $3.6$ divided by $\hbar\omega$ (in electronvolts), the pulse is physical and fulfills Maxwell equations.} Expressed for different wavelengths of light: the $\tau$ parameter should be bigger than 2~fs in the visible range and larger than 1.2~fs for UV pulses. Note that for a 2~fs pulse, the spectral bandwidth $\Omega$ is roughly 1 eV, and for a 1.2~fs pulse it is about 1.5 eV. Such bandwidths are broad enough to encompass the typical absorption bands of a molecule, making the vertical excitation approximation well justified for such short pulses. 

We note that we can easily define pulses fulfilling the condition of Eq.~(\ref{eq:cond1}) by using the vector potential $\Vec{A}(t)=\Vec{A}_0 A(t)$, where $\Vec{A}_0$ is the vector potential amplitude $A_0$ multiplied by the polarization vector $\Vec{\lambda}$. The electric field is defined from the vector potential as 
\begin{equation}
    \Vec{E}(t) = - \frac{1}{c}\frac{\dd \Vec{A}(t)}{\dd t} = - \frac{1}{c} \Vec{A}_0 \frac{\dd A(t)}{\dd t} = \Vec{A}_0 E(t) \, .
\end{equation}
Thus, any definition of $A(t)$ that gives a value of zero at the limits of $t=+\infty$ and $t=-\infty$ will create an electric field fulfilling the condition of Eq.~(\ref{eq:cond1}):
\begin{equation}
    \int_{-\infty}^\infty E(t) \dd t =  - \frac{1}{c} \int_{-\infty}^\infty \frac{\dd A(t)}{\dd t} \dd t = - \frac{1}{c} \left[ \lim_{t\to\infty}A(t) - \lim_{t\to-\infty}A(t) \right] = 0 \, .
\end{equation}

Returning to pulses defined by a pulse envelope, one can consider a vector potential in the following form 
\begin{equation}
    A(t) = - \frac{c}{\omega}\varepsilon(t) \sin(\omega t) \, .
\end{equation}
The scalar electric field obtained from such a definition reads
\begin{equation}
    E(t) = - \frac{1}{c}\frac{\dd A(t)}{\dd t} = \varepsilon(t) \cos(\omega t) + \frac{1}{\omega}\Dot{\varepsilon}(t) \sin(\omega t)  \, ,
\end{equation}
where the first term, $\varepsilon(t) \cos(\omega t)$, is the standard envelope times oscillating phase (real in this case) while the second term, $\frac{1}{\omega}\Dot{\varepsilon}(t) \sin(\omega t)$, acts like a correction ensuring that the condition of Eq.~(\ref{eq:cond1}) is fulfilled. Hence, we can define even ultrashort pulses in terms of envelopes, as long as we do so for the vector potential and work with the corresponding correction term. To connect the present discussion to the topics of the previous chapter, one would have to calculate the Wigner transformation from the full electric field formalism, i.e., calculate $\mathcal{W}_E$, and not just from the pulse envelope formalism ($\mathcal{W}_\varepsilon$). Finally, the equations above connect our work with the work of  Martínez-Mesa and Saalfrank\cite{Martinez-Mesa2015} who derived their equation for excited-state density in terms of $\mathcal{W}_\varepsilon$.

\clearpage

\section{Derivation of the time-dependent excited-state density}
\label{sec:deriv}

In this Section, we outline a complete derivation of the time-dependent excited-state density formula as presented in Eq. (8) in the main article. The original fundamental work on this topic was published by Li, Fang, and Martens\cite{Li1996} creating a framework for calculating pump-probe spectra. The work was then extended by Shen and Cina\cite{Shen1999} who alleviated some of the approximations introduced previously. While both works derived their leading formula in the density matrix formalism using the quantum Liouville equation, a similar formula was derived by Meier and Engel\cite{Meier2002} in 2002 from a wavefunction perspective rather than the density matrix perspective, avoiding problems with continuum states. Yet, all the aforementioned works considered a specific form of the laser envelope, usually a Gaussian one. The generalization to an arbitrary pulse envelope was achieved by Martínez-Mesa and Saalfrank,\cite{Martinez-Mesa2015} who realized that the interaction can be expressed as the pulse envelope Wigner transform $\mathcal{W}_\varepsilon$. Although this generalization brings some flexibility, it still relies on representing the pulse as an envelope times an oscillating phase, which may not always be justifiable for ultrashort pulses (see Section~\ref{sec:pulserep}). Furthermore, none of the works discussed above-derived equations considering the dependence of the transition dipole moment on nuclear positions (but instead considered it as constant). 
In the following paragraphs, we derive an equation for the excited-state density following a path similar to Meier and Engel, yet for a general electric field. Inspired by Martínez-Mesa and Saalfrank, we express the laser pulse in the Wigner representation, but without any restriction on its form -- making our formulation compatible with ultrashort pulses. We will show how the pulse envelope formulation appears from our formalism, offering a connection with the work of Martínez-Mesa and Saalfrank. We also generalize the formalism to position-dependent transition dipole moments. We finally propose a detailed discussion on the underlying assumptions of the strategy and the validity of its approximations. 

Let us consider a molecular system with two electronic states characterized by the (stationary) electronic wavefunctions $\phi_g$ (ground state) and $\phi_e$ (excited state), as well as time-dependent ground-state ($\psi_g(t)$) and excited-state ($\psi_e(t)$) nuclear wavefunctions. The total nuclear wavefunction for the system reads
\begin{equation}
    |\Psi (t)\rangle = 
    \begin{pmatrix}
        \psi_g(t) \\
        \psi_e(t)
    \end{pmatrix} \, .
\end{equation}
The Hamiltonian for our system, expressed in the basis of the two electronic states, can be separated into a time-independent Hamiltonian $\mathbf{H}_0$ and a small interaction term $\mathbf{V}_\mathrm{int}$ as is typical in perturbation theory, 
\begin{equation}
    \mathbf{H} = \mathbf{H}_0 + \mathbf{V}_\mathrm{int} = 
    \begin{pmatrix}
        \hat{H}_g &   0   \\
        0   &   \hat{H}_e
    \end{pmatrix} + 
    \begin{pmatrix}
        0   &   \hat{V}_\mathrm{int}  \\
        \hat{V}_\mathrm{int}  &   0   
    \end{pmatrix} \, .
\end{equation}
The components of the time-independent Hamiltonian,
\begin{align}
    \hat{H}_{g}(\mvec{R}) &= \hat{T} + E^\mathrm{el}_{g}(\mvec{R})\\
    \hat{H}_{e}(\mvec{R}) &= \hat{T} + E^\mathrm{el}_{e}(\mvec{R}) \, ,
\end{align}
depend on the nuclear kinetic energy operator $\hat{T}$ and the electronic energies $E^\mathrm{el}_{g/e}(\mvec{R})$, where $\mvec{R}$ is the the nuclear position vector. Note that the derivation is performed in the position representation; therefore, operators depending only on $\mvec{R}$ do not bear the $\wedge$ in our notation. The time-dependent interaction of the molecule with the electric field is defined within the dipole approximation as
\begin{equation}
    \hat{V}_\mathrm{int}(\mvec{R}, t) = -\Vec{\mu}_{eg}(\mvec{R})\cdot\Vec{E}_0 E(t) \, ,
    \label{eq:interaction}
\end{equation}
with $\Vec{\mu}_{eg}(\mvec{R})$ denoting the position-dependent transition dipole moment $\langle \phi_e|\hat{\mu} |\phi_g\rangle$, $\Vec{E}_0$ standing for the laser electric field amplitude, and $E(t)$ being a real time-dependent electric field. From now on, we will not specifically emphasize the dependence of $\hat{H}_{g/e}$, $E^\mathrm{el}_{g/e}$, $\hat{V}_\mathrm{int}$, and $\Vec{\mu}_{eg}$ on the nuclear positions $\mvec{R}$ for the sake of simplicity, yet it needs to be borne in mind during the derivation.

We want to solve the time-dependent Schrödinger equation,
\begin{equation}
    i\hbar\frac{\dd |\Psi (t)\rangle }{\dd t} = \left( \mathbf{H}_0 + \mathbf{V}_\mathrm{int} \right) |\Psi (t)\rangle  \, ,
\end{equation}
in terms of first-order perturbation theory, where the nuclear wavefunction is defined through the zero-order and first-order wavefunctions $|\Psi\rangle  = |\Psi^{(0)}\rangle +|\Psi^{(1)}\rangle $. Considering the initial conditions $\psi_g^{(1)}(t_0) = \psi_e^{(1)}(t_0) = \psi_e^{(0)}(t_0) = 0$, and $\psi_g^{(0)}(t_0) = \psi_g$, the working equations read
\begin{align}
    |\Psi^{(0)}(t)\rangle  &= \mathrm{e}^{-\frac{i}{\hbar}\mathbf{H}_0 (t-t_0)}|\Psi (t_0)\rangle \\
    |\Psi^{(1)}(t)\rangle  &= -\frac{i}{\hbar} \mathrm{e}^{-\frac{i}{\hbar}\mathbf{H}_0 t} \int_{t_0}^{t} \mathrm{e}^{\frac{i}{\hbar}\mathbf{H}_0 t^\prime} \mathbf{V}_\mathrm{int} (t^\prime)  |\Psi^{(0)}(t^\prime)\rangle  \dd t^\prime \, .\label{eq:first_pt}
\end{align}
Solving the equations for the excited-state nuclear wavefunction leads to
\begin{equation}
\label{eq:psi_pt}
    \psi_e(t) = -\frac{i}{\hbar}\mathrm{e}^{-\frac{i}{\hbar}\hat{H}_e t} \int_{t_0}^{t} \mathrm{e}^{\frac{i}{\hbar}\hat{H}_e t^\prime} \hat{V}_\mathrm{int}(t^\prime) \mathrm{e}^{-\frac{i}{\hbar}\hat{H}_g (t^\prime-t_0)} \psi_g \dd t^\prime \, .
\end{equation}
Reading the equation from the right, the initial ground-state nuclear wavefunction $\psi_g$ is propagated in the ground electronic state from time $t_0$ to $t^\prime$, at time $t^\prime$ it interacts with the laser pulse and gets promoted to the excited electronic state, before being (back) propagated to time $0$.\footnote{To clarify, the propagation does not go from $t^\prime$ back to time $t_0$ but to the time zero ($t=0$). Nevertheless, we could backpropagate to an arbitrary time given we would then propagate from this arbitrary time back to the time $t$. In the end, the operator acting is $\mathrm{e}^{-\frac{i}{\hbar}\hat{H}_e (t-t^\prime)}$. For convenience in our derivation, we work with time 0.} This is done for all times between $t$ and $t_0$ yielding what we will call a time-zero excited-state wavefunction $\psi_{z}$,
\begin{equation}
    \psi_{z}(t) = \int_{t_0}^{t} \mathrm{e}^{\frac{i}{\hbar}\hat{H}_e t^\prime} \hat{V}_\mathrm{int} \mathrm{e}^{-\frac{i}{\hbar}\hat{H}_g (t^\prime-t_0)} \psi_g \dd t^\prime \, .
    \label{eq:psiz}
\end{equation}
The wavefunction $\psi_{z}$ is constructed such that, if propagated in the excited state from time $0$ to time $t$, it yields the correct excited-state wavefunction $\psi_e(t)$,
\begin{equation}
    \psi_e(t) = -\frac{i}{\hbar}\mathrm{e}^{-\frac{i}{\hbar}\hat{H}_e t} \psi_{z} \, .
    \label{eq:psi_prop}
\end{equation}
The wavefunction $\psi_{z}$ is generally time-dependent since the upper integration limit depends on $t$. However, if time $t$ is a time after the pulse, it becomes a stationary (time-independent) wavefunction. Therefore, $\psi_{z}$ can be viewed as the initial condition already incorporating the laser pulse that only needs to be propagated in the excited state.
However, to obtain a useful formulation for trajectory-based techniques, we need to articulate this equation in terms of densities rather than wavefunctions.

\begin{figure}[ht!]
    \centering
    \includegraphics[width=1.0\linewidth]{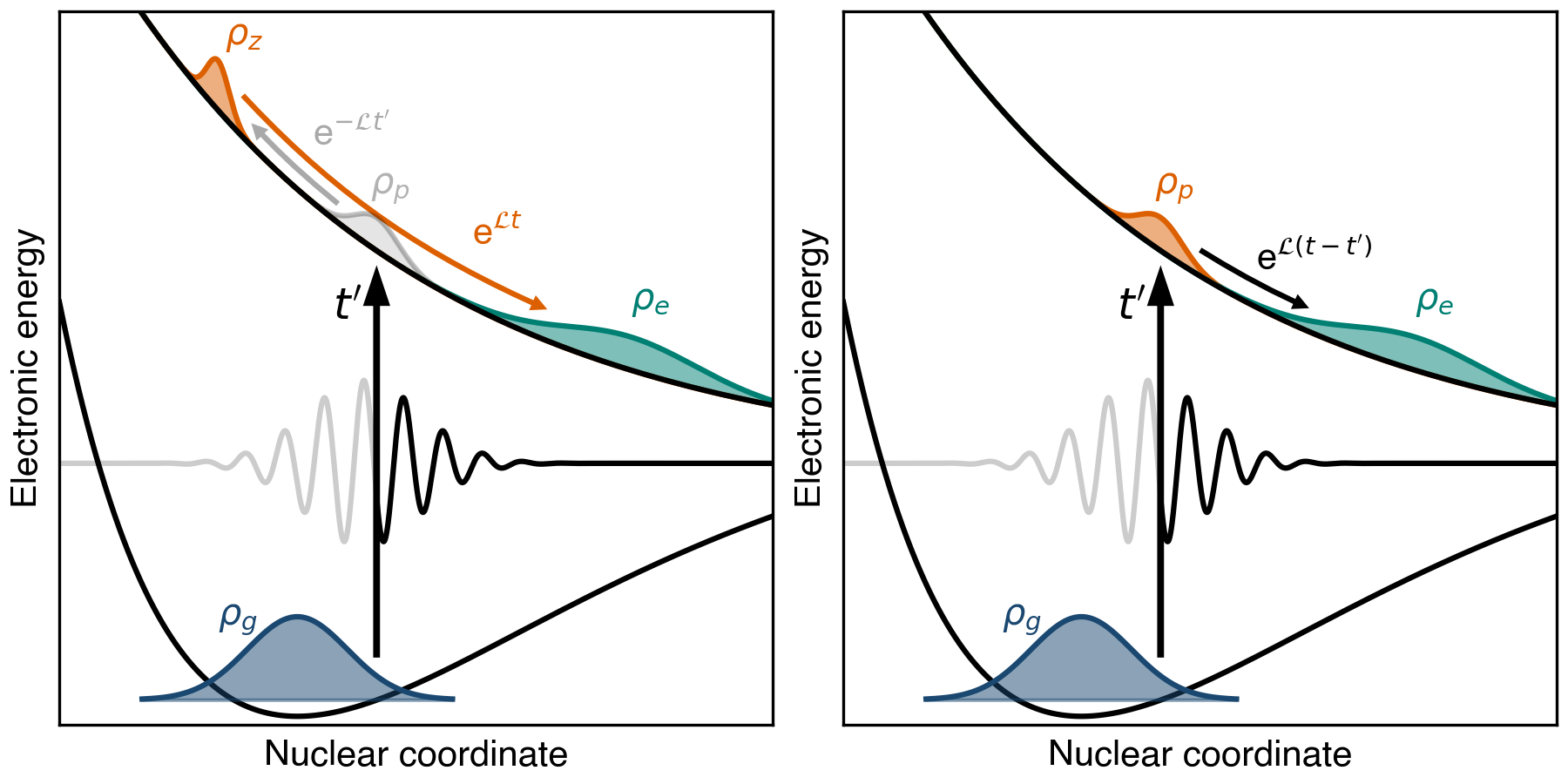}
    \caption{Schematic representation of the photoexcitation process and subsequent excited-state dynamics expressed in terms of the time-zero density $\rho_z$ and the promoted density $\rho_p$ (left), as well as only the promoted density $\rho_p$ (right). Both pictures lead to the same excited-state density $\rho_e$.}
    \label{fig:rhozpillust}
\end{figure}

The excited-state density $\rho_e(t) = |\psi_e(t)\rangle \langle \psi_e(t)|$ can be written in terms of time-zero excited-state density $\rho_{z}(t) = |\psi_{z}\rangle \langle \psi_{z}|$ and the quantum Liouville operator in the excited state $\mathcal{L}_e$,
\begin{equation}
    \rho_e (t) = |\psi_e\rangle \langle \psi_e| = \frac{1}{\hbar^2}\mathrm{e}^{-\frac{i}{\hbar}\hat{H}_e t} |\psi_{z} \rangle \langle \psi_{z} | \mathrm{e}^{\frac{i}{\hbar}\hat{H}_e t} = \frac{1}{\hbar^2} \mathrm{e}^{\mathcal{L}_e t} \rho_{z}(t) \, .
    \label{eq:dens_prop}
\end{equation}
From now on, we focus only on the dynamics after the laser pulse. Thus, we set the integration limits in Eq.~(\ref{eq:psiz}) such that they encompass the whole pulse, making $\rho_z$ time-independent. Since the interaction is zero when the laser pulse is over, we can integrate from $-\infty$ to $\infty$ without any loss of generality for dynamics after the pulse. Eq.~\eqref{eq:dens_prop} is then valid only for times $t$ after the laser pulse. The time-independent density $\rho_{z}$ now takes the form
\begin{align}
    \rho_{z} &= \int_{-\infty}^{\infty} \int_{-\infty}^{\infty}  \mathrm{e}^{\frac{i}{\hbar}\hat{H}_e \tau^\prime} \hat{V}_\mathrm{int}(\tau^\prime) \mathrm{e}^{-\frac{i}{\hbar}\hat{H}_g \tau^\prime} |\psi_g\rangle \langle \psi_g| \mathrm{e}^{\frac{i}{\hbar}\hat{H}_g \tau} \hat{V}_\mathrm{int}^*(\tau)  \mathrm{e}^{-\frac{i}{\hbar}\hat{H}_e \tau}\dd \tau^\prime  \dd \tau \, .
    \label{eq:promden0}
\end{align}
We now consider that the initial ground-state density $\rho_g$ is an eigenfunction of the time-independent ground-state Hamiltonian $\hat{H}_g$.\footnote{This means that $\rho_g$ is stationary under $\hat{H}_g$ and that $\mathrm{e}^{-\frac{i}{\hbar}\hat{H}_g t^\prime} |\psi_g\rangle = \mathrm{e}^{-\frac{i}{\hbar}E^0_g t^\prime} |\psi_g\rangle$.} 
Substituting $\tau = t^\prime + \frac{s}{2}$ and $\tau^\prime = t^\prime - \frac{s}{2}$, we get 
\begin{align}
    \rho_{z} &= \int_{-\infty}^{\infty} \mathrm{e}^{\frac{i}{\hbar}\hat{H}_e t^\prime}  \left[\int_{-\infty}^{\infty}  \mathrm{e}^{-\frac{i}{\hbar}\hat{H}_e \frac{s}{2}}  \hat{V}_\mathrm{int}\left(t^\prime - \frac{s}{2}\right) \mathrm{e}^{\frac{i}{\hbar}\hat{H}_g\frac{s}{2}} \rho_g \mathrm{e}^{\frac{i}{\hbar}\hat{H}_g \frac{s}{2}}  \hat{V}^*_\mathrm{int}\left(t^\prime + \frac{s}{2}\right) \mathrm{e}^{-\frac{i}{\hbar}\hat{H}_e \frac{s}{2}} \dd s \right] \mathrm{e}^{-\frac{i}{\hbar}\hat{H}_e t^\prime} \dd t^\prime \\
    &= \int_{-\infty}^{\infty} \mathrm{e}^{\frac{i}{\hbar}\hat{H}_e t^\prime}  \rho_p(t^\prime) \mathrm{e}^{-\frac{i}{\hbar}\hat{H}_e t^\prime} \dd t^\prime
    = \int_{-\infty}^{\infty} \mathrm{e}^{-\mathcal{L}_e t^\prime} \rho_p(t^\prime) \dd t^\prime \, , \label{eq:promden1}
\end{align}
where $\mathrm{e}^{-\mathcal{L}_e t^\prime}$ is a backward propagator from time $t^\prime$ to time 0 and what we denoted as a promoted density $\rho_p$ is defined as 
\begin{align}
    \rho_p(t^\prime) &=  \int_{-\infty}^{\infty}  \mathrm{e}^{-\frac{i}{\hbar}\hat{H}_e \frac{s}{2}}  \hat{V}_\mathrm{int}\left(t^\prime - \frac{s}{2}\right) \mathrm{e}^{\frac{i}{\hbar}\hat{H}_g\frac{s}{2}} \rho_g \mathrm{e}^{\frac{i}{\hbar}\hat{H}_g \frac{s}{2}}  \hat{V}^*_\mathrm{int}\left(t^\prime + \frac{s}{2}\right) \mathrm{e}^{-\frac{i}{\hbar}\hat{H}_e \frac{s}{2}} \dd s \, .
    \label{eq:rhof1}
\end{align}
The distinction between $\rho_z$ and $\rho_p$ and their interpretation are represented graphically 
in Fig.~\ref{fig:rhozpillust}. 

Substituting the definition of interaction potential (Eq.~\eqref{eq:interaction}) into Eq.~\eqref{eq:rhof1}, we get
\begin{align}
    \rho_p(t^\prime) &=  \int_{-\infty}^{\infty} E\left(t^\prime + \frac{s}{2}\right) E\left(t^\prime - \frac{s}{2}\right) \mathrm{e}^{-\frac{i}{\hbar}\hat{H}_e \frac{s}{2}} \left[\Vec{\mu}_{eg}\cdot\Vec{E}_0\right] \mathrm{e}^{\frac{i}{\hbar}\hat{H}_g\frac{s}{2}} \rho_g \mathrm{e}^{\frac{i}{\hbar}\hat{H}_g \frac{s}{2}} \left[\Vec{\mu}_{eg}\cdot\Vec{E}_0\right] \mathrm{e}^{-\frac{i}{\hbar}\hat{H}_e \frac{s}{2}} \dd s \, ,
    \label{eq:rhof2}
\end{align}
where we consider the electric field as a real function, i.e., $E^*(t)=E(t)$. The task is now to deal with the operator $\mathrm{e}^{-\frac{i}{\hbar}\hat{H}_e \frac{s}{2}} \left[\Vec{\mu}_{eg}\cdot\Vec{E}_0\right] \mathrm{e}^{\frac{i}{\hbar}\hat{H}_g\frac{s}{2}}$. If we consider that the transition dipole moment is a constant, we can take advantage of the Baker--Campbell--Hausdorff (BCH) formula ($\mathrm{e}^{\hat{A}}\mathrm{e}^{\hat{B}} = \mathrm{e}^{\hat{A}+\hat{B}+\frac{1}{2}[\hat{A},\hat{B}]+\cdots}$) to its first order, and express the operator as an exponential of the electronic energy difference
\begin{equation}
    \mathrm{e}^{-\frac{i}{\hbar}\hat{H}_e \frac{s}{2}} \left[\Vec{\mu}_{eg}\cdot\Vec{E}_0\right] \mathrm{e}^{\frac{i}{\hbar}\hat{H}_g\frac{s}{2}} = \Vec{\mu}_{eg}\cdot\Vec{E}_0  \mathrm{e}^{-\frac{i}{\hbar}\hat{H}_e \frac{s}{2}} \mathrm{e}^{\frac{i}{\hbar}\hat{H}_g\frac{s}{2}} \stackrel{\mathrm{BCH}}{\approx} \Vec{\mu}_{eg}\cdot\Vec{E}_0 \mathrm{e}^{-\frac{i}{\hbar}(E^\mathrm{el}_e - E^\mathrm{el}_g)  \frac{s}{2}} \, .
\end{equation}
While this approach is convenient, we decided here to retain the dependence of the transition dipole moment on the nuclear coordinates and instead to apply the first-order Baker--Campbell--Hausdorff formula twice
\begin{align}
        \mathrm{e}^{-\frac{i}{\hbar}\hat{H}_e \frac{s}{2}} \left[\Vec{\mu}_{eg}\cdot\Vec{E}_0\right] \mathrm{e}^{\frac{i}{\hbar}\hat{H}_g\frac{s}{2}} &= \mathrm{e}^{-\frac{i}{\hbar}\hat{H}_e \frac{s}{2}} \mathrm{e}^{\ln(\Vec{\mu}_{eg}\cdot\Vec{E}_0)} \mathrm{e}^{\frac{i}{\hbar}\hat{H}_g\frac{s}{2}} \\
        &\stackrel{\mathrm{BCH}}{\approx} \mathrm{e}^{-\frac{i}{\hbar}\hat{H}_e \frac{s}{2} + \ln(\Vec{\mu}_{eg}\cdot\Vec{E}_0)} \mathrm{e}^{\frac{i}{\hbar}\hat{H}_g\frac{s}{2}} \\
        &\stackrel{\mathrm{BCH}}{\approx} \mathrm{e}^{-\frac{i}{\hbar}(\hat{H}_e - \hat{H}_g)\frac{s}{2} + \ln(\Vec{\mu}_{eg}\cdot\Vec{E}_0)} \\
        &= \mathrm{e}^{-\frac{i}{\hbar}(E^\mathrm{el}_e - E^\mathrm{el}_g)  \frac{s}{2}+ \ln(\Vec{\mu}_{eg}\cdot\Vec{E}_0)} = \Vec{\mu}_{eg}\cdot\Vec{E}_0 \mathrm{e}^{-\frac{i}{\hbar}\Delta E^\mathrm{el}_{eg}  \frac{s}{2}} \label{eq:bch1} \, .
\end{align}
These approximations are in general valid for short times (meaning that the laser pulse should be of a short duration) and slowly varying $\Vec{\mu}_{eg}$ and $\Delta E^\mathrm{el}_{eg}$. The limits of these approximations will be studied at the end of this Section, with all the other approximations and assumptions made throughout the derivation. Taking Eq.~\eqref{eq:bch1} and inserting it into Eq.~\eqref{eq:rhof2} leads to 
\begin{align}
    \rho_p(t^\prime) &= |\Vec{\mu}_{eg}\cdot\Vec{E}_0|^2 \int_{-\infty}^{\infty} E(t^\prime + \frac{s}{2}) E(t^\prime - \frac{s}{2}) \mathrm{e}^{-\frac{i}{\hbar}\Delta E^\mathrm{el}_{eg} s} \dd s \rho_g \\
    &= |\Vec{\mu}_{eg}\cdot\Vec{E}_0|^2 \mathcal{W}_E(t^\prime,\Delta E^\mathrm{el}_{eg}/\hbar) \rho_g \, , \label{eq:finalden}
\end{align}
where we have identified the Wigner representation of the pulse $\mathcal{W}_E$ as defined in Eq.~\eqref{eq:def1}. Our Eq.~\eqref{eq:finalden} for a general form of laser pulses can be also written in terms of the pulse envelope Wigner transform $\mathcal{W}_\varepsilon$, as introduced by Martínez-Mesa and Saalfrank\cite{Martinez-Mesa2015}, using the identity defined in Eq.~\eqref{eq:wigtowig2}, leading to 
\begin{align}
    \rho_p(t^\prime) = |\Vec{\mu}_{eg}\cdot\Vec{E}_0|^2  \mathcal{W}_\varepsilon(t^\prime,\Delta E^\mathrm{el}_{eg}/\hbar-\Dot{\gamma}) \rho_g \, , 
    \label{eq:promden}
\end{align}
which is designed for laser pulses defined as an envelope times an oscillating phase (see Eq.~\eqref{eq:pulse}).

The time-dependent excited-state density $\rho_e$ from Eq.~(\ref{eq:dens_prop}), combined with Eq.~(\ref{eq:promden1}) and Eq.~(\ref{eq:finalden}), now reads
\begin{align}
    \rho_e(t) &= \frac{1}{\hbar^2}  \int_{-\infty}^{\infty} \mathrm{e}^{\mathcal{L}_e (t-t^\prime)} \rho_p(t^\prime) \dd t^\prime \notag \\
    &= \frac{1}{\hbar^2}  \int_{-\infty}^{\infty} \mathrm{e}^{\mathcal{L}_e (t-t^\prime)} \left[|\Vec{\mu}_{eg}\cdot\Vec{E}_0|^2 \mathcal{W}_E(t^\prime,\Delta E^\mathrm{el}_{eg}/\hbar)  \rho_g  \right] \dd t^\prime
    \label{eq:dens_prop2}
\end{align}
and provides a clear interpretation of the photoexcitation process. For each time $t^\prime$, the stationary ground-state density is multiplied by the Wigner pulse transform at time $t^\prime$ and by the squared projection of the transition dipole moment on the electric field amplitude $|\Vec{\mu}_{eg}\cdot\Vec{E}_0|^2$ -- we denote this the promoted density $\rho_p$. The density promoted at time $t^\prime$ is then propagated in the excited electronic state from time $t^\prime$ until the desired time $t$.  Finally, integration over all the times $t^\prime$ reconstructs the excited-state density at time $t$. 

\begin{figure}[ht!]
    \centering
    \includegraphics[width=1\linewidth]{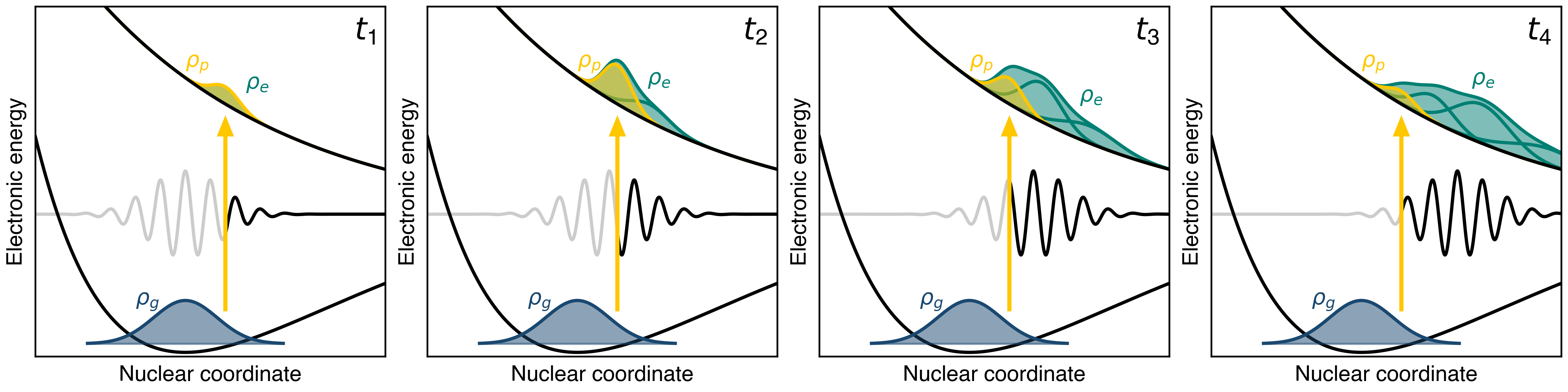}
    \caption{Illustration of the promoted density approach for photoexcitation and subsequent dynamics. The promoted density is depicted at four different times ($t_1<t_2<t_3<t_4$) along with the excited-state density, which consists of propagated promoted densities from earlier times (see also Fig.~\ref{fig:rhozpillust}).}
    \label{fig:rhopillust}
\end{figure}

As stated earlier, Eq.~\eqref{eq:dens_prop2} is formally valid only for times $t$ after the pulse duration since we changed the upper integration limit from $t$ to $\infty$ in Eq.~\eqref{eq:psiz}. This modification was necessary to obtain the Wigner pulse transform $\mathcal{W}_E$, which requires integration from $-\infty$ to $\infty$, in the equations. However, this modification leads to a loss of validity during and before the pulse. To clarify this statement, let us examine the behavior of Eq.~\eqref{eq:dens_prop2} during the pulse interaction by splitting the integral into two as
\begin{equation}
    \rho_e(t) = \frac{1}{\hbar^2}  \left( \int_{-\infty}^{t} \mathrm{e}^{\mathcal{L}_e (t-t^\prime)} \rho_p(t^\prime) \dd t^\prime  + \int_{t}^{\infty} \mathrm{e}^{\mathcal{L}_e (t-t^\prime)} \rho_p(t^\prime) \dd t^\prime \right) \, .
    \label{eq:notepda3}
\end{equation}
The first integral takes the promoted density $\rho_p$ excited before the current time ($t^\prime < t$) and propagates it forward in time with $\mathrm{e}^{\mathcal{L}_e (t-t^\prime)}$. Conversely, the second integral goes over excitation times larger than the current time ($t^\prime > t$) and, therefore, takes the promoted density in the \textit{future} and propagates it backward in time with $\mathrm{e}^{- \mathcal{L}_e |t-t^\prime|}$ as $t-t^\prime < 0$. In other words, the first integral accounts for the density that has been promoted to the excited state, while the second integral takes care of the density that is yet to be promoted to the excited state. Thus, if our current time $t$ is after the pulse, the second integral is equal to zero and we can write
\begin{align}
    \rho_e(t) &= \frac{1}{\hbar^2}  \int_{-\infty}^{t} \mathrm{e}^{\mathcal{L}_e (t-t^\prime)} \rho_p(t^\prime) \dd t^\prime \notag \\
    &= \frac{1}{\hbar^2}  \int_{-\infty}^{t} \mathrm{e}^{\mathcal{L}_e (t-t^\prime)} \left[|\Vec{\mu}_{eg}\cdot\Vec{E}_0|^2 \mathcal{W}_E(t^\prime,\Delta E^\mathrm{el}_{eg}/\hbar)  \rho_g  \right] \dd t^\prime \, . \label{eq:dens_prop2b}
\end{align}
If our current time $t$ is during or before the pulse, neglecting the second integral means that we discard the \textit{future} excited-state density and only account for the part that has been already excited. We retain only the first integral in PDA, trying to alleviate the restriction of Eq.~\eqref{eq:dens_prop2} to describe only the density for times after the pulse.
The excitation picture stemming from Eq.~\eqref{eq:dens_prop2b} is illustrated in Fig.~\ref{fig:rhopillust}, and is easy to convert into an algorithm for modeling photoexcitation processes in pump-probe excitation, see Section~\ref{sec:alg}. 

The final approximation to make this scheme applicable to trajectory-based nonadiabatic dynamics approaches is to take the Wigner representation of the density operators, retaining only the lowest order of $\hbar$. The classical excited-state density then reads
\begin{equation}
\boxed{
    \rho^\mathrm{cl}_e(\mvec{R},\mvec{P},t) = \frac{1}{\hbar^2} \int_{-\infty}^{t} \mathrm{e}^{\mathcal{L}^\mathrm{cl}_e (t-t^\prime)} \left[ |\Vec{\mu}_{eg}(\mvec{R})\cdot\Vec{E}_0|^2 \mathcal{W}_E(t^\prime,\Delta E^\mathrm{el}_{eg}(\mvec{R})/\hbar)  \rho^\mathrm{cl}_g(\mvec{R},\mvec{P}) \right] \dd t^\prime \, .
}
    \label{eq:dens_prop3}
\end{equation}
We emphasized the dependence of all the terms on nuclear positions $\mvec{R}$ and momenta $\mvec{P}$ for clarity. Eq.~\eqref{eq:dens_prop3} is the final equation for simulating excited-state dynamics triggered by a laser pulse. The scheme emanating from Eq.~\eqref{eq:dens_prop3} is simple: one only needs to calculate the promoted density $\rho_p$ for all times $t'$ during the pulse and then propagate this promoted density in the excited electronic states with the standard techniques of nonadiabatic dynamics, e.g., trajectory surface hopping or \textit{ab initio} multiple spawning. Eq.~\eqref{eq:dens_prop3} also provides a simple recipe to construct the promoted density $\rho_p$: take the ground-state density and multiply it by the Wigner pulse representation and the squared projection of the transition dipole moment on the electric field amplitude.
 
Let us finish this Section by summarizing the different assumptions and approximations made to derive Eq.~\eqref{eq:dens_prop3}. 
\begin{enumerate}
    \item Eq.~\eqref{eq:dens_prop3} was derived within the framework of first-order perturbation theory with all its assumptions, meaning that this equation cannot be used for two-photon processes, strong-field regimes, etc.
    \item The off-diagonal matrix elements, like the nonadiabatic coupling terms, were considered to be zero in our starting Hamiltonian. This assumption decouples the photoexcitation process from the nonadiabatic ones and might not be well justified if the nonadiabatic couplings are non-negligible in the Franck-Condon region. 
    \item PDA was initially derived to describe the nuclear density in the excited state after the laser pulse, yet we attempted to alleviate this approximation in Eq.~\eqref{eq:dens_prop2b}, without affecting the obtained excited-state density after the pulse. This modification improves the dynamics during the laser pulse (see Figs.~\ref{fig:longpulse} and \ref{fig:lorentzzoom}), however, it should be taken with care.
    \item The derivation of Eq.~\eqref{eq:dens_prop3} assumes a position-dependent transition dipole moment, utilizing the Baker--Campbell--Hausdorff formula to first order. To assess the validity of this assumption, we discuss here the importance of the second-order term. In the following, we will use the notation $\Tilde{\mu} = \Vec{\mu}_{eg}\cdot\Vec{E}_0 $ for simplicity. The second-order Baker--Campbell--Hausdorff formula reads
    \begin{align}
        \mathrm{e}^{-\frac{i}{\hbar}\hat{H}_e \frac{s}{2}} \Tilde{\mu}  &= \mathrm{e}^{-\frac{i}{\hbar}\hat{H}_e \frac{s}{2}} \mathrm{e}^{\ln(\Tilde{\mu})} \stackrel{\mathrm{BCH}}{\approx} \mathrm{e}^{-\frac{i}{\hbar}\hat{H}_e \frac{s}{2} + \ln(\Tilde{\mu}) - \frac{is}{4\hbar}\left[\hat{T},\ln(\Tilde{\mu})\right]}  \, ,
    \end{align}
    where the commutator takes the form
    \begin{equation}
        \left[\hat{T},\ln(\Tilde{\mu})\right] = -\frac{i\hbar}{2m}\left( \hat{p}\frac{\dd \ln(\Tilde{\mu})}{\dd \hat{x}} + \frac{\dd \ln(\Tilde{\mu})}{\dd \hat{x}}\hat{p}\right) = \frac{\hbar^2}{2m}\left[ \left(\frac{\Tilde{\mu}^\prime}{\Tilde{\mu}}\right)^2 - \frac{\Tilde{\mu}^{\prime\prime}}{\Tilde{\mu}} - 2\frac{\Tilde{\mu}^\prime}{\Tilde{\mu}} \frac{\dd}{\dd x} \right] \, .
    \end{equation}    
    Hence, the first correction to the transition dipole moment will be small if (i) the derivatives of the transition dipole moment vary slowly compared to its value and (ii) the laser pulses are short as we integrate the term over the pulse duration. The approximation might not be justified for long laser pulses and quickly varying transition dipole moments.
    \item We have also used a short-time approximation by using the first-order Baker--Campbell--Hausdorff formula for the propagator $\mathrm{e}^{-\frac{i}{\hbar}\hat{H}_e \frac{s}{2}}  \mathrm{e}^{\frac{i}{\hbar}\hat{H}_g\frac{s}{2}}$. This approximation can again be relaxed by considering the second-order term 
    \begin{equation}
        \mathrm{e}^{-\frac{i}{\hbar}\hat{H}_e \frac{s}{2}}  \mathrm{e}^{\frac{i}{\hbar}\hat{H}_g\frac{s}{2}} \stackrel{\mathrm{BCH}}{\approx}  \mathrm{e}^{-\frac{i}{\hbar}\Delta E^\mathrm{el} _{eg}\frac{s}{2} - \frac{1}{8\hbar^2}\left[\hat{T},\Delta E^\mathrm{el} _{eg}\right]  s^2} \, ,
    \end{equation}
    where the commutator can be written as
    \begin{equation}
        \left[\hat{T},\Delta E^\mathrm{el} _{eg}\right] = -\frac{i\hbar}{2m}\left( \hat{p}\frac{\dd \Delta E^\mathrm{el} _{eg}}{\dd \hat{x}} + \frac{\dd \Delta E^\mathrm{el} _{eg}}{\dd \hat{x}}\hat{p}\right) = -\frac{\hbar^2}{2m}\left( \frac{\dd^2 \Delta E^\mathrm{el}_{eg}}{\dd x^2} + 2\frac{\dd \Delta E^\mathrm{el} _{eg}}{\dd x}\frac{\dd}{\dd x}\right) \, .
    \label{eq:secordcont}
    \end{equation}
    Combining these two expressions together results in
    \begin{equation}
        \mathrm{e}^{-\frac{i}{\hbar}\hat{H}_e \frac{s}{2}}  \mathrm{e}^{\frac{i}{\hbar}\hat{H}_g\frac{s}{2}} \stackrel{\mathrm{BCH}}{\approx}  \mathrm{e}^{-\frac{i}{\hbar}\Delta E^\mathrm{el} _{eg}\frac{s}{2} + \frac{1}{16m}\left( \frac{\dd^2 \Delta E^\mathrm{el}_{eg}}{\dd x^2} + 2\frac{\dd \Delta E^\mathrm{el} _{eg}}{\dd x}\frac{\dd}{\dd x}\right)  s^2} \, .
    \end{equation}
    As stressed in point 4 above, this analysis means that Eq.~\eqref{eq:dens_prop3} might not be justified for long laser pulses and significant nuclear gradient differences. 
    
    Let us try to establish a range of validity to neglect the second-order contribution. If one retains only the first term on the right-hand side of Eq.~\eqref{eq:secordcont} (as the derivative makes the exponential difficult to evaluate), the pulse duration should follow the following condition: 
    \begin{equation}
        \frac{1}{8m} \frac{\dd^2 \Delta E^\mathrm{el}_{eg}}{\dd x^2} \tau^2 < 1 \, .
    \end{equation}
    Note that the factor 8 instead of 16 appears because we have the exponential twice in the derivation. Considering the photoexcitation of NaI from its ground-state minimum, we obtain $\tau < 45$~fs.
    \item The transition to classical nuclear densities assumed the truncation of the terms with a higher order in $\hbar$ in the exact Wigner representation of the quantum operators. The approach is then not suitable for situations where strong quantum effects play a role (which is compatible in any case with the use of most trajectory-based approaches to nonadiabatic dynamics). As such, we expect PDA to be valid for molecular cases where the nuclear ensemble method is capable of adequately describing absorption spectra.
\end{enumerate}

\clearpage
\section{Algorithm to use the promoted density approach for single and multiple excited states}
\label{sec:alg}

In the main text, we have proposed a practical implementation of Eq.~\eqref{eq:dens_prop3}, coined promoted density approach (PDA). We provide here its algorithmic description (Algorithm~\ref{alg:v1}). The algorithm randomly selects the (nuclear) position-momentum pairs \{$\mvec{R}_i$,~$\mvec{P}_i$\} from the (approximate) ground-state density, randomly selects an excitation time $t^\prime$, and calculates a probability $p$ based on Eq.~\eqref{eq:dens_prop3}. At that stage, negative probabilities must be handled if they occur. The probability $p$ is then compared to a random number generated from a uniform distribution in the range 0 to $p_\mathrm{max}$, where $p_\mathrm{max}$ is the maximum probability that can be determined from the available position-momentum pairs and excitation times.

\begin{algorithm}
\caption{The PDA algorithm used to sample the promoted density $\rho^\mathrm{cl}_p(\mvec{R},\mvec{P},t^\prime)$. The input consists of the ground-state nuclear position-momentum pairs \{$\mvec{R}_i$,~$\mvec{P}_i$\} with their corresponding excitation energies $\Delta E_i = \Delta E_{eg}^\mathrm{el}(\mvec{R}_i)$ and transition dipole moments $\Vec{\mu}_i = \Vec{\mu}_{eg}(\mvec{R}_i)$.}\label{alg:v1}
estimate maximum probability $p_\mathrm{max}$ \\
$j=1$\\
\While{$j \leq N_p$}
{
    randomly select $i\in\{1,\dots,N_g\}$\\
    randomly select $t^\prime$\\
    calculate probability $p = |\Vec{\mu}_i\cdot\Vec{E}_0|^2\mathcal{W}_E(t^\prime, \Delta E_i)$\\
    \If{$p < 0$}
    {
        handle negative probabilities (see Section~\ref{sec:negprob})
    }
    randomly select $\mathcal{R} \in [0, p_\mathrm{max}]$\\
    \If{$\mathcal{R} \leq p$}
    {
        accept \{$\mvec{R}_i$,~$\mvec{P}_i$,~$t^\prime$\} as an initial condition $j$\\
        $j=j+1$
    }
}
\end{algorithm}

The algorithm above can be easily extended for multiple excited states. Considering more excited states in the derivation proposed in Section~\ref{sec:deriv} means that first-order perturbation theory would provide us with the same uncoupled formula (Eq.~\eqref{eq:psi_pt}) for each excited electronic state. Thus, we would end up with an Eq.~\eqref{eq:dens_prop3} for each electronic state considered. Hence, extending PDA to multiple electronic states requires only one additional step: randomly selecting the excited state $s$ from a set of $N_s$ excited states, see Algorithm~\ref{alg:v2}. Note that in such a case, the initial conditions generated by PDA contain additional information about the excited state in which the nonadiabatic dynamics should be initiated, i.e., \{$\mvec{R}_j$,~$\mvec{P}_j$,~$t^\prime_j$,~$s_j$\}.

\begin{algorithm}
\caption{The PDA algorithm used to sample the promoted densities $\rho^{\mathrm{cl},s}_p(\mvec{R},\mvec{P},t^\prime)$ considering multiple excited electronic states denoted by the index $s$. The input consists of the ground-state nuclear position-momentum pairs \{$\mvec{R}_i$,~$\mvec{P}_i$\} with their corresponding excitation energies $\Delta E_i^s = \Delta E_{sg}^\mathrm{el}(\mvec{R}_i)$ and transition dipole moments $\Vec{\mu}_i^s = \Vec{\mu}_{sg}(\mvec{R}_i)$ from the ground state $g$ to one of the $N_s$ excited electronic states considered.}\label{alg:v2}
estimate maximum probability $p_\mathrm{max}$ \\
$j=1$\\
\While{$j \leq N_p$ }
{
    randomly select $i\in\{1,\dots,N_g\}$\\
    randomly select $s\in\{1,\dots,N_s\}$\\
    randomly select $t^\prime$\\
    calculate probability $p = |\Vec{\mu}_i^s\cdot\Vec{E}_0|^2\mathcal{W}_E(t^\prime, \Delta E_i^s)$\\
    \If{$p < 0$}
    {
        handle negative probabilities (see Section.~\ref{sec:negprob})
    }
    randomly select $\mathcal{R} \in [0, p_\mathrm{max}]$\\
    \If{$\mathcal{R} \leq p$}
    {
        accept \{$\mvec{R}_i$,~$\mvec{P}_i$,~$t^\prime$,~$s$\} as an initial condition $j$\\
        $j=j+1$
    }
}
\end{algorithm}

\clearpage

\section{Computational details}

\subsection{NaI -- sodium iodide}

\begin{figure}[ht!]
    \centering
    \includegraphics[width=0.7\textwidth]{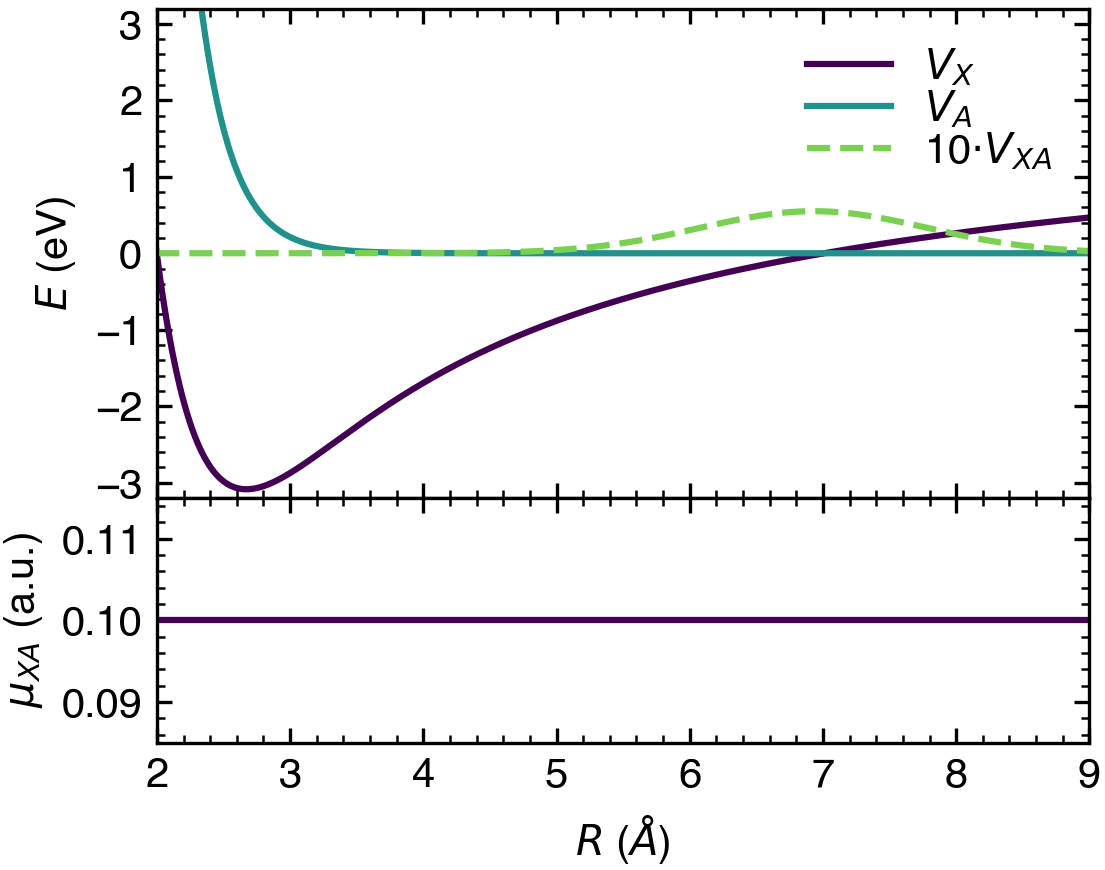}
    \caption{A visual representation of the terms used to build the NaI diabatic Hamiltonian defined in Eq.~(\ref{eq:diabham}).}
    \label{fig:nai_ham}
\end{figure}

The reference (numerically-exact) quantum dynamics simulations were performed with the split-operator technique in the diabatic basis using a time step of 0.25 a.u. and on a grid with 8192 points between 3.7 and 75 a.u. The convergence of the results with respect to the grid and the time step was thoroughly tested. The NaI diabatic Hamiltonian was reproduced from Ref.~\citenum{Martinez-Mesa2015} (originally coming from Ref.~\citenum{Engel1989}) and is defined as
\begin{equation}
    \mathbf{H}_\mathrm{d} = \hat{T}\mathbf{1}_2 +
    \begin{pmatrix}
        V_{X} & V_{XA} + V_\mathrm{int}(t)\\
        V_{XA} + V_\mathrm{int}(t) & V_{A}
    \end{pmatrix} \, , 
    \label{eq:diabham}
\end{equation}
where $\hat{T}$ is the kinetic energy operator, $V_{X}$ and $V_{A}$ are the diabatic potential energy curves for the ionic state $X(^1\Sigma^+)$ and covalent state $A(0^+)$ respectively, $V_{XA}$ is the diabatic coupling, and $V_\mathrm{int}$ is the interaction term defined as
\begin{equation}
    V_\mathrm{int}(t) = -\Vec{\mu}_{XA}\cdot\Vec{E}_0 E(t) = -\mu_{XA}E_0 E(t)\, ,
    \label{eq:interaction2}
\end{equation}
where $\mu_{XA}$ and $E_0$ are magnitudes of their respective vectors. The diabatic potential energy curves and the diabatic coupling are expressed in the following form:
\begin{align}
    V_X(R) &= \left[ A_2 + \left( \frac{B_2}{R}\right)^8\right] \exp\left(-\frac{R}{\rho}\right) - \frac{e^2}{R} - \frac{e^2\left(\lambda^+ + \lambda^-\right)}{2R^4} - \frac{C_2}{R^6} - \frac{2e^2\lambda^+\lambda^-}{R^7} + \Delta E_0 \, , \\
    V_A(R) &= A_1 \exp\left[-\beta_1 (R-R_0)\right] \, , \\
    V_{XA}(R) &= A_{12} \exp\left[-\beta_{12} (R-R_x)^2\right] \, ,
\end{align}
with all parameters summarized in Tab.~\ref{tab:hparams}. A visual representation of the Hamiltonian is provided in Fig.~\ref{fig:nai_ham}.

\begin{table}[ht!]
    \centering
    \begin{tabular}{c l}
    \hline
       parameter  &  value \\
       \hline
       $A_2$  &        2760 eV \\
       $B_2$  &        2.398 eV$^{1/8}$ \AA\\
       $C_2$  &        11.3 eV \AA$^{6}$ \\
       $\lambda^+$  &  0.408 \AA$^3$ \\
       $\lambda^-$  &  6.431 \AA$^3$ \\
       $\rho$  &       0.3489 \AA \\
       $\Delta E_0$  & 2.075 eV \\
       $e^2$  &        14.3996 eV \AA \\
       \hline
       $A_1$  &        0.813 eV \\
       $\beta_1$  &    4.08 \AA$^{-1}$  \\
       $R_0$  &        2.67 \AA \\
       \hline
       $A_{12}$  &     0.055 eV \\
       $\beta_{12}$  & 0.6931 \AA$^{-2}$  \\
       $R_x$  &        6.93 \AA \\
       \hline
       $\mu_{XA}$  &  0.1 a.u.\\
       \hline
       $m$ & 35480.251398 a.u. \\
       \hline
       \end{tabular}
    \caption{Parameters defining the NaI diabatic Hamiltonian, reproduced from Ref.~\citenum{Martinez-Mesa2015} (the original diabatic parameters come from Ref.~\citenum{Engel1989}). The mass $m$ was taken as the reduced mass of NaI.}
    \label{tab:hparams}
\end{table}

The electric field intensity $E_0$ was set to 0.001 a.u. to remain within the weak-field regime, resulting in a maximum of 0.05\% population transfer for the longest pulse. The scalar electric field $E(t)$ is defined in Eq.~(11) in the main text.
The precise pulse frequencies used in the work are $\omega_0=0.15250790$, $0.14294844$, and $0.13520905$~a.u.

The bare diabatic Hamiltonian, that is, without the $V_\mathrm{int}(t)$ term,  was transformed into the adiabatic representation for the FSSH dynamics through the following transformation matrix
\begin{equation}
    \mathbf{U} = 
    \begin{pmatrix}
        \cos\theta & \sin\theta \\
        -\sin\theta & \cos\theta
    \end{pmatrix} \, ,
\end{equation}
where $\theta$ is the mixing angle defined as
\begin{equation}
    \theta = \frac{1}{2}\arctan\frac{2V_{XA}}{V_{XX}-V_{AA}} \, .
\end{equation}
The adiabatic potential energy curves then read
\begin{equation}
    E^\mathrm{el}_{1,2} = \frac{V_{XX}+V_{AA}}{2}\pm\frac{1}{2}\sqrt{(V_{AA}-V_{XX})^2+4V_{XA}^2}
\end{equation}
and the nonadiabatic coupling vector is defined as 
\begin{equation}
    d_{12} = \langle \phi_1^\mathrm{ad}| \frac{d}{dR} \phi_2^\mathrm{ad}\rangle = - \frac{d\theta}{dR} \, .
\end{equation}

The FSSH simulations were performed with an energy-decoherence parameter 0.1 a.u.\cite{Granucci2007} and a time step of 2.5 a.u. using the molecular dynamics code ABIN\cite{Hollas2019}. We note that the previous study of NaI by Martínez-Mesa and Saalfrank\cite{Martinez-Mesa2015} reported only a minor effect of the decoherence parameter on the predissociation dynamics. 10'000 position-momentum pairs were sampled from the ground-state Wigner distribution (the harmonic approximation was not invoked) and propagated with the FSSH method. We stress again here that the FSSH simulations \textit{do not include the explicit interaction with a laser pulse}.

\subsection{Protonated formaldimine}

To simulate the photodynamics of protonated formaldimine, we reused 500 FSSH trajectories from our previous work on this molecule.\cite{Suchan2020} These 145fs-long trajectories employed a time step of 0.24 fs and an energy-decoherence correction with the decoherence parameter 0.1 a.u.\cite{Granucci2007} The electronic structure was described by the FOMO-CASCI method,\cite{Slavicek2010} considering 12 electrons in 8 orbital and a 6-31G* basis set. The Gaussian broadening parameter was set to 0.2 a.u. More details about the methodology are available in the original Ref.~\citenum{Suchan2020}.

\subsection{Calculation of observables}

Calculating observables with PDA is a slightly more complex task than when the vertical sudden excitation is invoked. While all the (FSSH) trajectories are initiated in an excited state at the very same time within the sudden vertical excitation, the number of trajectories in the excited state increases gradually within PDA. This observation leads to an important question: \textit{how should one evaluate an observable before and during the pulse when not all the trajectories have yet been promoted?}

We propose in this work to consider that each trajectory starts at time $-\infty$ and remains 'fixed' (or frozen) at their ground-state geometry until the excitation time $t^\prime$ is reached and the trajectory gets promoted to the excited state, initiating its evolution. This strategy is based on the Eq.~(\ref{eq:dens_prop3}), where the system is initially described by its stationary ground-state density $\rho_g$ before chunks of it get promoted to the excited state. We shall illustrate this approach for the specific case of electronic-state populations. We consider the molecule in its ground state ($g$) until time $t^\prime$ when it gets promoted to the excited state. From that time, the population is governed by the nonadiabatic dynamics (starting in the excited state), in other words, by the standard population $p_i^\mathrm{nonad}$ evaluated during typical FSSH or AIMS simulations. Hence, the time-dependent electronic state populations take the form:
\begin{equation}
    p_i(t) = 
    \begin{cases}
        g & t < t^\prime \\
        p_i^\mathrm{nonad}(t) & t>t^\prime \, .
    \end{cases} 
\end{equation}

Conversely, evaluating observables in a specific electronic state ($s$) is simply based on trajectories active in that state at any given moment, i.e.,
\begin{equation}
    \mathcal{O}^s(t) = \frac{\sum_{i=1}^{N^s(t)} \mathcal{O}_i^s(t)}{N^s(t)} \, ,
    \label{eq:observe}
\end{equation}
where $\mathcal{O}^s$ is the desired observable in the electronic state $s$, $N^s(t)$ is the time-dependent number of trajectories propagated in the state $s$, and $\mathcal{O}_i^s$ is the observable evaluated for the trajectories $i$. The sum is taken over all trajectories propagating in the given state $s$ at time $t$.

\subsubsection{Population transfer in QD and PDA}

While the comparison of $\langle R\rangle_{S_1}$ and $\langle \Delta R\rangle_{S_1}$ between QD and FSSH+PDA(W) can be done directly from the simulation data due to the presence of a normalization in the formula for an expectation value, the comparison of electronic populations shown in this Supporting Information requires a normalization of the QD populations. As we mentioned at the beginning of this Section, the field intensity $E_0$ was set to trigger a maximum of 0.05\% population transfer to the excited electronic state, ensuring that our simulations are in the weak-field limit. Contrarily, PDA considers only the promoted part of the density (0.05\%) and ignores the part remaining in the ground state. Thus, the QD electronic state populations must be normalized such that they reflect only the excited part of the density. In practice, we have taken the maximum of the QD excited-state population (0.05\%) and rescaled it to 1. Note that this rescaling strategy was only possible as the depopulation of the excited electronic state due to nonadiabatic transitions happens long after the pulse. 

\clearpage
\section{Extended tests of PDA and PDAW}
\vspace{-0.15cm}

In this Section, we present a series of extended tests of PDA and PDAW (as mentioned above and in the main article) based on the photodynamics of NaI.

\vspace{-0.35cm}
\subsection{Comparing PDA, PDAW, and standard windowing}

\vspace{-0.35cm}
\begin{figure}[ht!]
    \centering
    \includegraphics[width=0.93\textwidth]{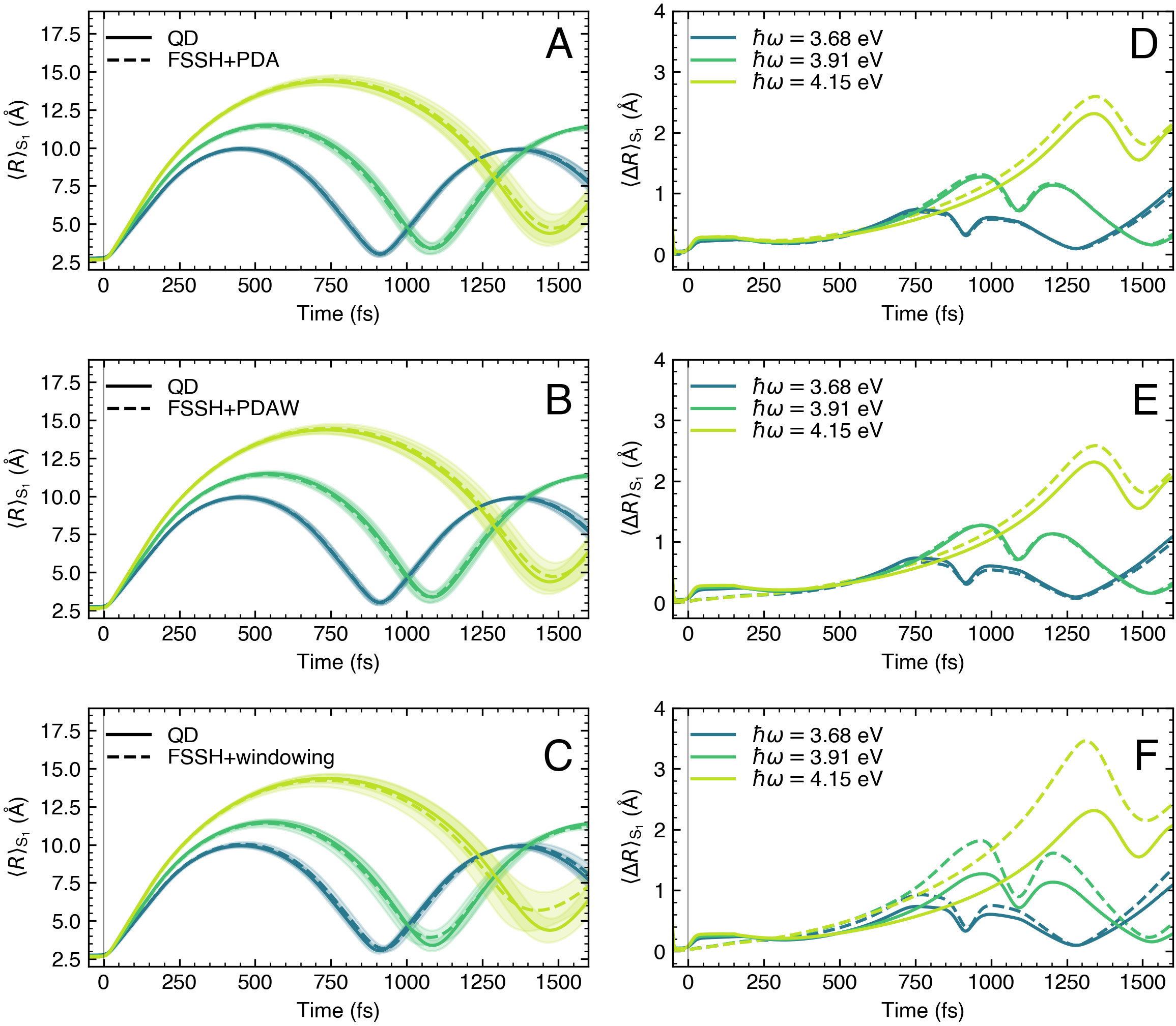}
    \caption{Supplement to Fig.~3 from the main article, including a comparison of FSSH simulations using PDAW. The results highlight the improved performance of the windowing strategy (PDAW) with respect to the most commonly used windowing approach. FSSH+PDAW matches the results of FSSH+PDA. (A) Expectation values of the NaI bond length in $S_1$ for three different laser pulse frequencies, comparing quantum dynamics with an explicit 20-fs laser pulse (solid lines) and PDA combined with FSSH nonadiabatic dynamics (dashed lines). The shaded area represents $\langle \Delta R \rangle_{\mathrm{S}_1}$ of the nuclear wavepacket (QD) or trajectories (FSSH). (B) Same as in panel A but for FSSH combined with PDAW (dashed lines). (C) Same as in panel A but for FSSH using a simple windowing approach combined with a time convolution (dashed lines). (D) The width of the excited-state nuclear wavepacket $\langle \Delta R \rangle_{\mathrm{S}_1}$ corresponding to simulations in panel A (similar correspondence for panel E and panel F). }
    \label{fig:si_nai_omega}
\end{figure}

\clearpage
\subsection{Excitation with a chirped pulse: PDA vs. PDAW}

\begin{figure}[ht!]
    \centering
    \includegraphics[width=1\textwidth]{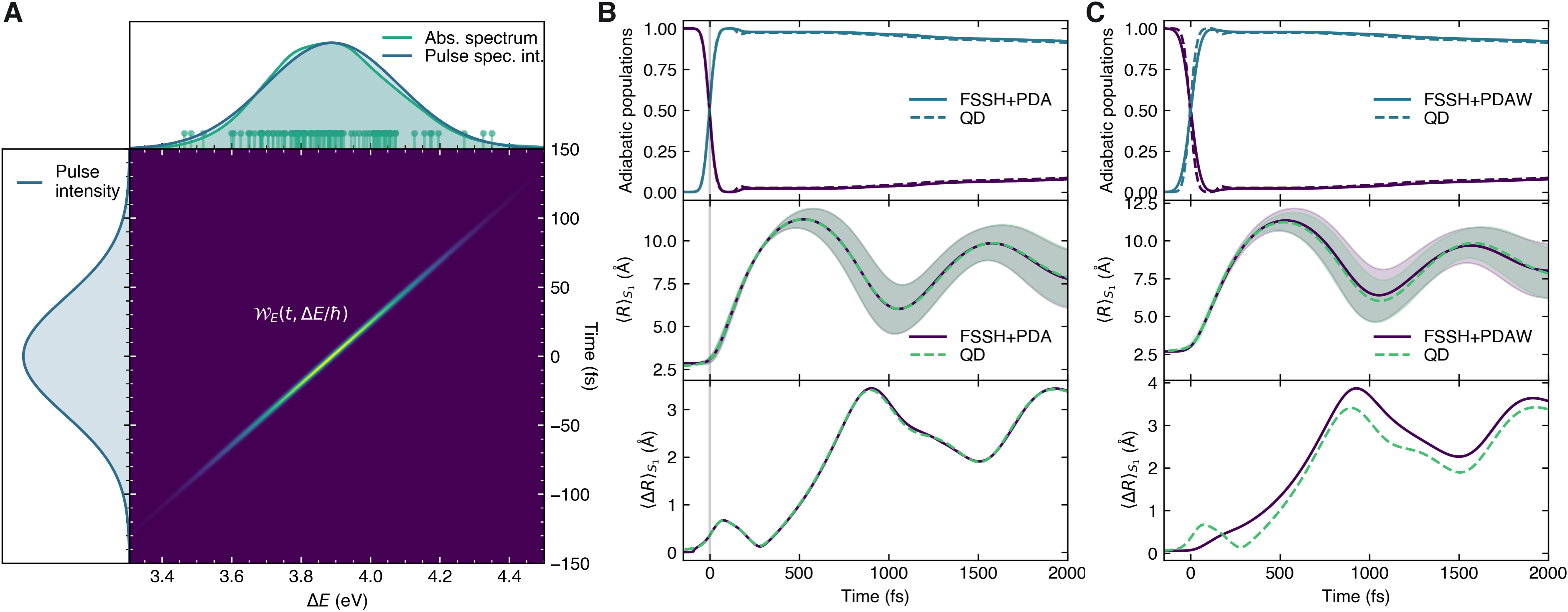}
    \caption{Photodynamics of NaI following excitation with a 100-fs Gaussian pulse with $\omega_0=0.14294844$~a.u. and a linear chirp parameter $\beta=2\times10^{-6}$~a.u. (A) The Wigner pulse representation of the chirped Gaussian pulse plotted with the absorption spectrum, pulse intensity, and spectral intensity. (B) Adiabatic electronic state populations, expectation values of the NaI bond length in $S_1$, and its standard deviation $\langle \Delta R \rangle_{\mathrm{S}_1}$ for PDA combined with FSSH nonadiabatic dynamics (dashed lines), compared to the reference QD with an explicit laser pulse (solid lines). (C) Same as in panel B but for PDAW. FSSH+PDA is in excellent agreement with the QD results and outperforms FSSH+PDAW. Although not quantitative, FSSH+PDAW can still capture the pulse effects qualitatively, demonstrating a weak effect of the chirp.}
    \label{fig:chirpedpulse}
\end{figure}

\clearpage
\subsection{Long laser pulses with PDA}

\begin{figure}[ht!]
    \centering
    \includegraphics[width=\textwidth]{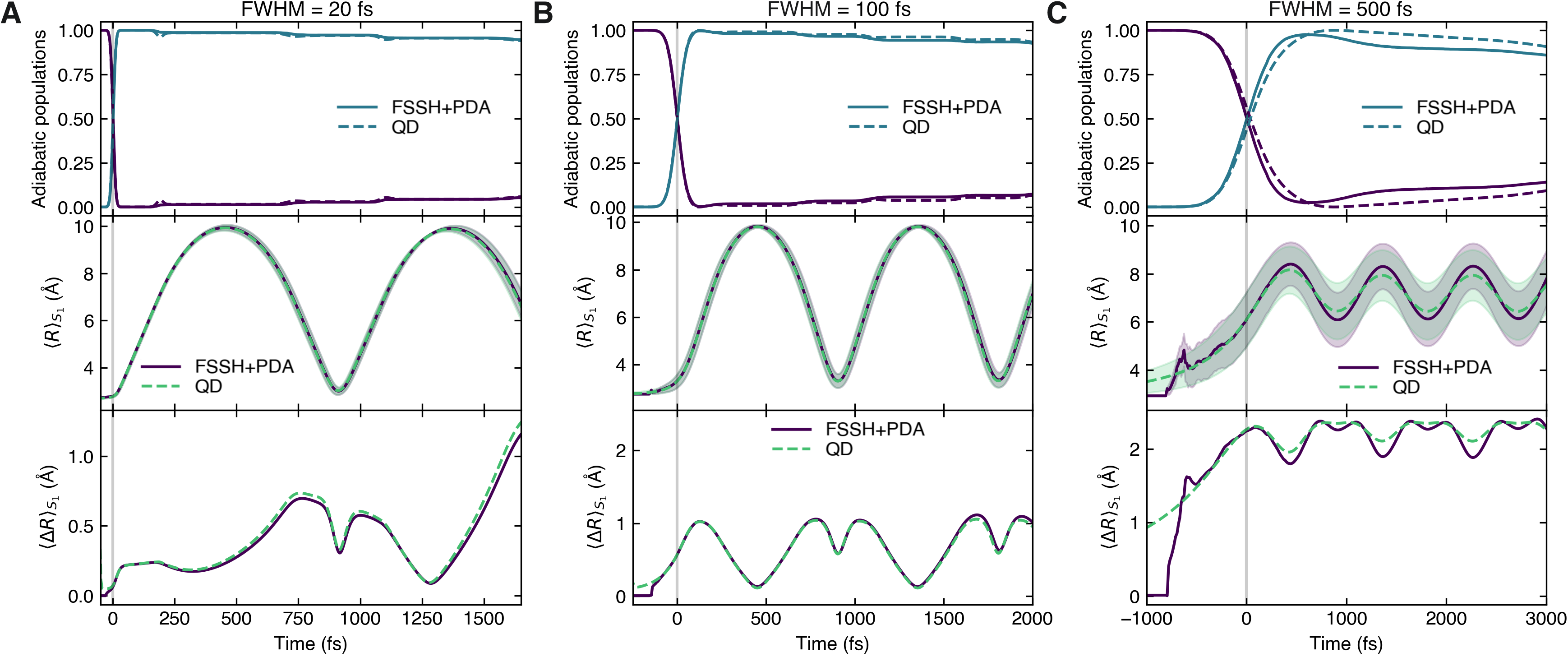}
    \caption{Photoexcitation of NaI triggered by a Gaussian laser pulse with a frequency $\omega_0=0.13520905$~a.u. and three different FWHM parameter $\tau$, comparing QD to FSSH combined with PDA. (A) Adiabatic electronic state populations, expectation values of the NaI bond length in $S_1$, and its standard deviation $\langle \Delta R \rangle_{\mathrm{S}_1}$ for a 20-fs laser pulse. PDA combined with FSSH nonadiabatic dynamics (dashed lines) is compared to the reference QD with an explicit pulse (solid lines). (B) Same as in panel A but for a 100-fs laser pulse. (C) Same as in panel A but for a 500-fs laser pulse.}
    \label{fig:longpulse}
\end{figure}

As discussed in Section~\ref{sec:deriv}, one of the approximations behind PDA assumes a short duration of the laser pulse. Based on the second-order Baker--Campbell--Hausdorff (BCH) formula, we have estimated (in the specific case of the photodynamics of NaI) that the pulse duration should be shorter than 45~fs for this approximation to be valid. So far, we have applied only pulses with a maximum value of $\tau=20$~fs, well below this estimated limit. Hence, we present here additional simulations with 100-fs and 500-fs laser pulses to explore the boundaries of the BCH approximation. The results presented in Fig.~\ref{fig:longpulse} reveal that the effect of the 100-fs laser pulse is still perfectly captured by PDA. For the 500-fs pulse, we observe a sizeable deviation from the QD result. However, we would rather attribute this deviation to quantum interferences within the excited nuclear wavepacket caused by the long laser pulse, since the period of oscillation in the excited state is similar to the pulse duration. Still, FSSH+PDA manages to depict the adiabatic nuclear dynamics qualitatively, while the time evolution of the adiabatic populations and the nuclear wavepacket width are better captured. These results indicate that PDA is more robust with respect to the laser pulse duration than one would expect from our estimated limit.

\clearpage
\subsection{Lorentzian envelope: negative Wigner probabilities and dynamics during the pulse}
\label{sec:negprob}

\begin{figure}[h!]
    \centering
    \includegraphics[width=0.70\textwidth]{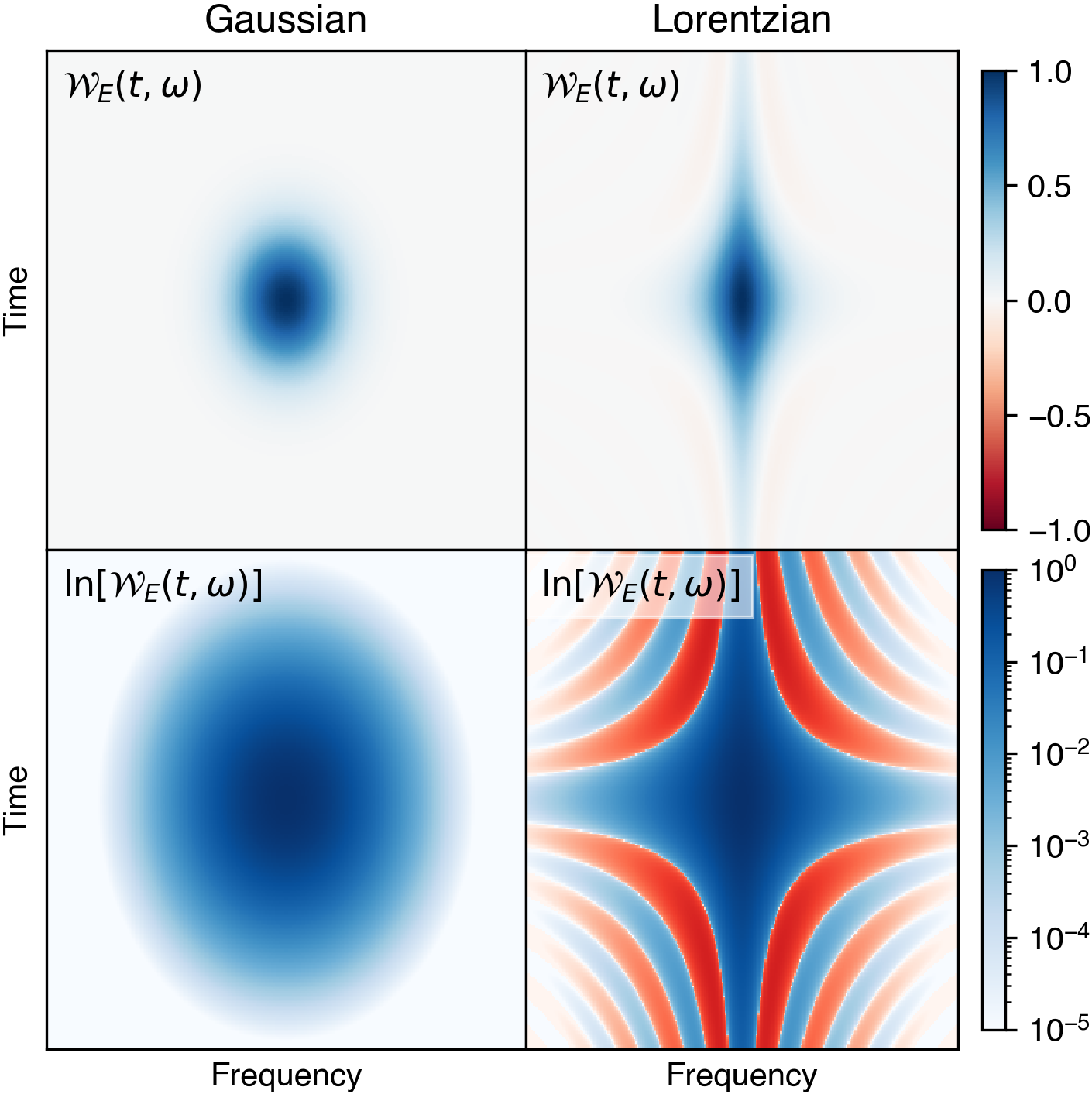}
    \caption{The Wigner representation of a Gaussian and a Lorentzian laser pulse. Blue depicts positive values of the corresponding function, while red highlights its negative parts. The top panels show $\mathcal{W}_E$ using a linear scale, demonstrating the small contribution of negative values to the overall $\mathcal{W}_E$. The bottom panels show $\mathcal{W}_E$ in a logarithmic scale, which emphasizes the structure of negative regions (barely visible using a linear standard scale).}
    \label{fig:neglorentz}
\end{figure}

The use of a Gaussian laser pulse in the main text and previous Sections is an ideal case for PDA, since the Wigner representation of Gaussian pulses is strictly positive. However, we show in the following that PDA can also be combined with other pulse envelopes once a suitable strategy for dealing with negative probabilities stemming from $\mathcal{W}_E$ is introduced. Although $\mathcal{W}_E$ should not be seen as a probability function, the PDA scheme considers $\mathcal{W}_E$ to be a probability, leading to issues when negative values appear (for non-Gaussian laser pulses). First, let us define the Lorentzian envelope,
\begin{equation}
    \varepsilon(t) = \left[1 + \frac{4}{1 + \sqrt{2}}\left(\frac{t}{\tau}\right)^2\right]^{-1} \, ,
\end{equation}
where $\tau$ is the FWHM parameter for the intensity $I(t) \approx \varepsilon^2(t)$. The Wigner pulse representation of the Lorentzian envelope is shown in Fig.~\ref{fig:neglorentz}, highlighting its negative regions. The Gaussian envelope is also provided for comparison. Notice that the negative values emerge at the 'outskirts' of $\mathcal{W}_E$, with small magnitudes in comparison to the maximum of the function. As such, the negative regions appear as a minor contribution to $\mathcal{W}_E$ and it could be tempting to neglect them. 

\begin{figure}[ht!]
    \centering
    \includegraphics[width=1\textwidth]{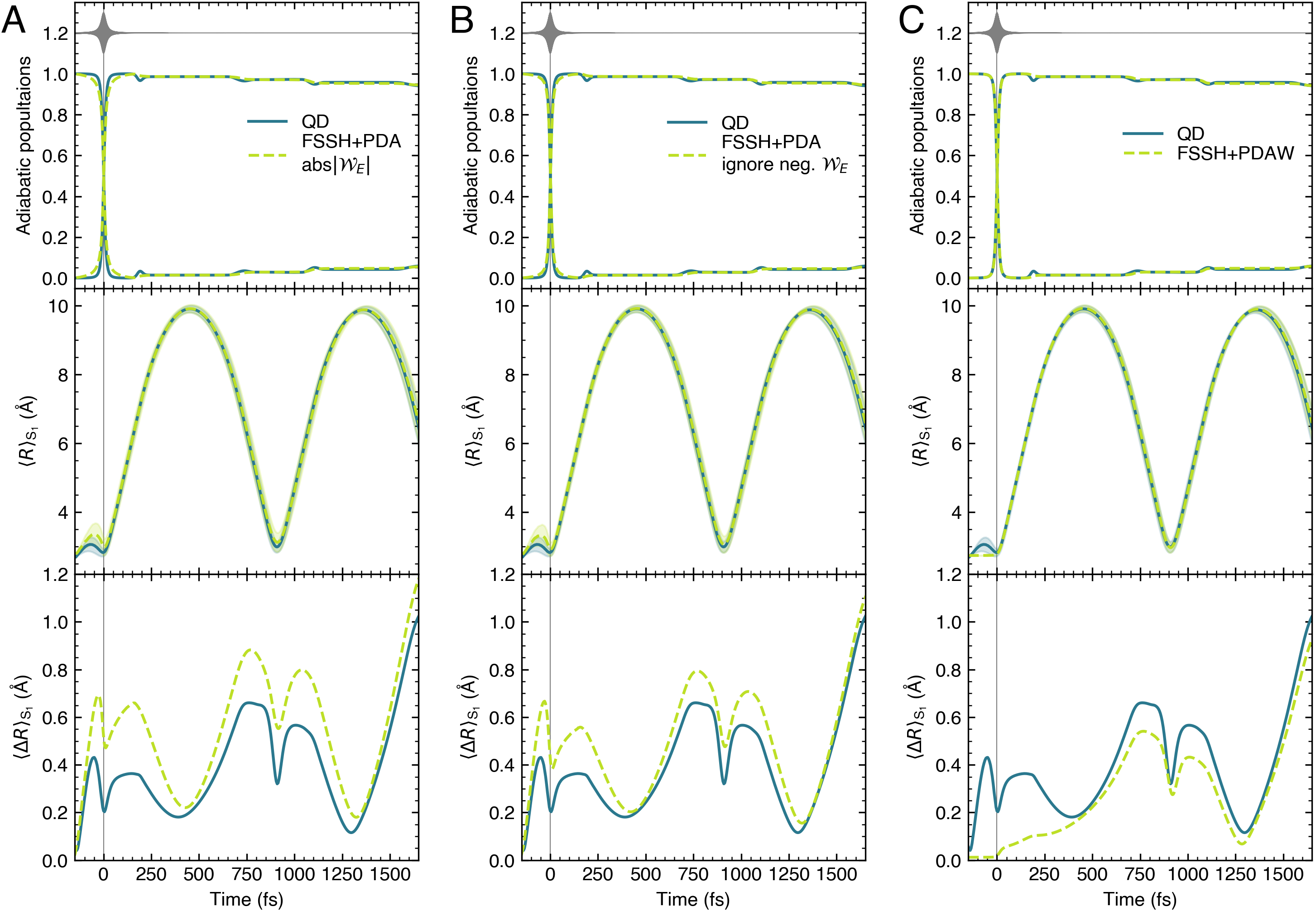}
    \caption{Photoexcitation of NaI with a Lorentzian laser pulse using a frequency $\omega_0=0.13520905$~a.u. and a FWHM parameter $\tau=20$~fs. The results of QD with an explicit laser pulse are compared to those of FSSH with either PDA or PDAW. (A) Comparing QD (solid lines) with FSSH combined with PDA (dashed lines), taking the absolute value of $\mathcal{W}_E$. The adiabatic populations, expectation values of the NaI bond length in $S_1$, and its standard deviation $\langle \Delta R \rangle_{\mathrm{S}_1}$ are reported. (B) Same as in panel A, but this time simply ignoring the negative values of $\mathcal{W}_E$, i.e., setting them to 0. (C) Same as in panel A, but using PDAW instead of PDA.}
    \label{fig:lorentz}
\end{figure}

\begin{figure}[ht!]
    \centering
    \includegraphics[width=1\textwidth]{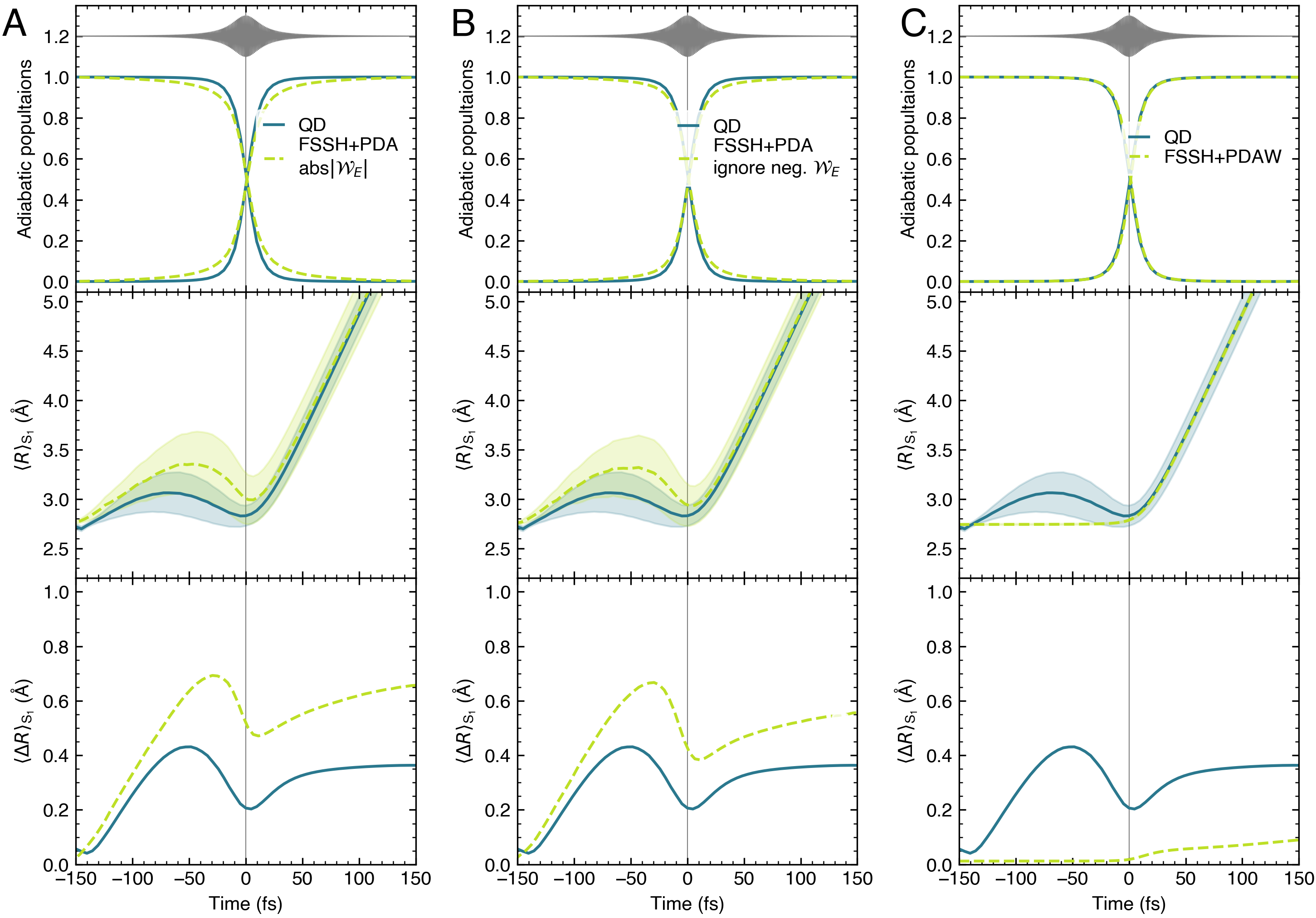}
    \caption{Zoom of the data reported in  Fig.~\ref{fig:lorentz} in the time window around the laser pulse (from -150~fs to 150~fs). The data demonstrate the ability of PDA to account for the nuclear dynamics also during the laser pulse (missed, as expected, by PDAW).}
    \label{fig:lorentzzoom}
\end{figure}

Based on this analysis, we propose two schemes for creating a modified Wigner pulse representation $\Tilde{\mathcal{W}}_E$ that can handle the negative values of $\mathcal{W}_E$. First, one can consider the absolute value of $\mathcal{W}_E$, i.e., $\Tilde{\mathcal{W}}_E = |\mathcal{W}_E|$. In this case, the negative regions contribute to $\mathcal{W}_E$ with small positive values, and the ground-state density can also be promoted to the excited state in these regions of $t$ and $\omega$. The second strategy consists of ignoring the negative values completely, i.e., substituting the negative values with zeros:
\begin{equation}
    \Tilde{\mathcal{W}}_E =
    \begin{cases}
        0  &\mathrm{if}\ \mathcal{W}_E(t,\omega)<0 \\
        \mathcal{W}_E(t,\omega) & \mathrm{elsewhere}
    \end{cases} 
\end{equation}
One could also consider PDAW as a strategy to handle negative values, since $I(t)$ and $S(\omega)$ are both positive functions. We tested all the aforementioned strategies on the photoexcitation of NaI triggered by a Lorentzian pulse with  $\omega_0=0.13520905$~a.u. and $\tau=20$~fs, see Fig.~\ref{fig:lorentz}. The results in Fig.~\ref{fig:lorentz} clearly demonstrate the ability of all these strategies to handle negative probabilities and capture the laser pulse effects at an almost quantitative level. The deviation from QD, observed for the populations and $\langle R \rangle_{\mathrm{S}_1}$, are nearly negligible. The only significant difference can be seen for the nuclear wavepacket width $\langle \Delta R \rangle_{\mathrm{S}_1}$, but it is still minor when compared to simulations using the sudden vertical excitation. Thus, all strategies to deal with negative values in the Wigner pulse representation appear to work for this test system, and we would favor neglecting the negative values based on the slightly better agreement in the nuclear wavepacket width between FSSH+PDA and the QD reference.

\clearpage

\section{Python implementation of PDA and PDAW}
\label{sec:implement}

Both PDA and PDAW were implemented in a user-friendly Python code called \verb|promdens|\cite{pda_code}, available to the public on \href{https://github.com/PHOTOX/promdens}{GitHub} or as a Python package via \href{https://pypi.org/project/promdens/}{PyPI}. The code generates the initial conditions (PDA) or weights and convolution parameters (PDAW) that can be used to run and process trajectory-based simulations from any nonadiabatic dynamics code. In this Section, we provide a brief set of guidelines on how to use this code, and we direct the interested reader to the \href{https://github.com/PHOTOX/promdens}{GitHub} repository for more details and the manual. 

\subsection{Installation} 
The code is published on \href{https://pypi.org/project/promdens/}{PyPI} and can be installed with \verb|pip|
\begin{lstlisting}[language=bash, columns=fullflexible,style=bash]
$ pip install promdens
\end{lstlisting}
After installation, the code is available as a script via the \bash{promdens} command. To print help, run:
\begin{lstlisting}[language=bash, columns=fullflexible,style=bash]
$ promdens --help
\end{lstlisting}
The minimum supported Python version is 3.7. The code depends on \verb|numpy| and \verb|matplotlib| libraries that are automatically installed by \verb|pip|. 

\subsection{Usage}

The code requires information about the method (PDA or PDAW), the number of excited states, the number of initials conditions to be generated, and the characteristics of the laser pulse, such as the envelope type (Gaussian, Lorentzian, sech, etc.), the pulse frequency, the linear chirp parameter, and the full width at half maximum parameter. The code can be launched from a terminal with a series of flags as follows
\begin{lstlisting}[language=bash, columns=fullflexible,style=bash]
$ promdens --method pda --energy_unit a.u. --tdm_unit debye --nstates 2 --fwhm 3 --omega 0.355 --npsamples 10 --envelope_type gauss input_file.dat
\end{lstlisting}

The input file should contain information about the excitation energies and magnitudes of the transition dipole moments for each pair of sampled nuclear positions and momenta (label by an index number).\footnote{If the user would like to consider the pulse polarization $\Vec{E}_0$ as well, the quantity $|\Vec{\mu}_{eg}\cdot\Vec{E}_0|$ should then be provided.}
In the following, we provide an example of the input file for the first two excited states of protonated formaldimine:
\begin{lstlisting}[language=bash, columns=fullflexible,style=bash]
#index    dE12 (a.u.)  |mu_12| (Debye)  dE13 (a.u.)  |mu_13| (Debye)
1         0.32479719       0.1251       0.40293672       1.351
2         0.32070472       0.2434       0.40915241       1.289
3         0.34574925       0.7532       0.38595754       1.209
4         0.33093699       0.1574       0.36679075       1.403
5         0.31860215       0.1414       0.36973886       1.377
6         0.31057768       0.0963       0.40031651       1.390
7         0.33431888       0.1511       0.40055704       1.358
8         0.31621589       0.0741       0.36644659       1.425
9         0.32905912       0.5865       0.36662982       1.277
10        0.31505412       0.2268       0.35529522       1.411
\end{lstlisting}

Using this input file and running the command line above, the user receives the following output file called \bash{pda.dat} containing information about excitation times and initial excited states:
\begin{lstlisting}[language=bash, columns=fullflexible,style=bash]
# Sampling: number of ICs = 10, number of unique ICs = 5
# Field parameters: omega = 3.55000e-01 a.u., linear_chirp = 0.00000e+00 a.u., fwhm = 3.000 fs, t0 = 0.000 fs, envelope type = 'gauss'
# index  exc. time (a.u.)  el. state     dE (a.u.)    |tdm| (a.u.)
     3     15.09731061          1       0.34574925     0.29635106
     3     25.94554064          1       0.34574925     0.29635106
     3     61.98106992          1       0.34574925     0.29635106
     4      7.38522206          2       0.36679075     0.55201877
     8    -14.27561557          2       0.36644659     0.56067480
     9    155.72500917          2       0.36662982     0.50244331
     9    -44.31379959          2       0.36662982     0.50244331
    10     94.19109952          2       0.35529522     0.55516642
    10     -9.13220842          2       0.35529522     0.55516642
    10     31.75086044          2       0.35529522     0.55516642
\end{lstlisting}
Inspecting this output file shows that the code generated 10 initial conditions accounting for the effect of the laser pulse, yet only 5 unique ground-state samples (pairs of nuclear positions and momenta) were used: indexes 3, 9, and 10 were selected more than once. The initial conditions are also spread over both excited states. The user should then run only 5 nonadiabatic simulations: initiating the nuclear position-momentum pair with index 3 in the first excited state and the nuclear position-momentum pairs with indexes 4, 8, 9, and 10 in the second excited state.

If the same command were to be used with PDAW instead of PDA (\bash{--method pdaw}), the output file would look as follows
\begin{lstlisting}[language=bash, columns=fullflexible,style=bash]
# Convolution: I(t) = exp(-4*ln(2)*(t-t0)^2/fwhm^2)
# Parameters:  fwhm = 3.000 fs, t0 = 0.000 fs
# index        weight S1        weight S2
       1      1.78475e-05      9.66345e-07
       2      1.56842e-05      2.59858e-08
       3      6.31027e-02      1.29205e-03
       4      1.79107e-04      1.62817e-01
       5      2.31817e-06      1.01665e-01
       6      2.96548e-08      3.90152e-06
       7      3.81650e-04      3.33694e-06
       8      2.36147e-07      1.75628e-01
       9      1.47188e-03      1.37747e-01
      10      1.33347e-06      3.55670e-01
\end{lstlisting}
The code provides the pulse intensity and weights necessary for the convolution described in Eq.~(15) in the main text. Note that the intensity should be normalized before being used in the convolution. If only a restricted number of trajectories can be calculated, the user should choose the indexes and initial excited states corresponding to the largest weights in the file. For example, if one could run only 10 trajectories for the protonated formaldimine, we would run the nuclear position-momentum pairs with indexes 3, 4, 7, and 9 starting in $S_1$ and indexes 3, 4, 5, 8, 9, and 10 starting in $S_2$.

If the user selects the option \bash{--plot}, the code will produce a series of plots analyzing the provided data and calculated results, e.g. the absorption spectrum calculated with the nuclear ensemble approach, the pulse spectrum, or the Wigner pulse transform.

More information about the code is available in the manual on the GitHub repository (the code is also thoroughly commented). 

\clearpage
%\bibliography{references}% Produces the bibliography via BibTeX.
% 
\providecommand{\latin}[1]{#1}
\makeatletter
\providecommand{\doi}
  {\begingroup\let\do\@makeother\dospecials
  \catcode`\{=1 \catcode`\}=2 \doi@aux}
\providecommand{\doi@aux}[1]{\endgroup\texttt{#1}}
\makeatother
\providecommand*\mcitethebibliography{\thebibliography}
\csname @ifundefined\endcsname{endmcitethebibliography}
  {\let\endmcitethebibliography\endthebibliography}{}